\catcode`\@=11
\newif\if@fewtab\@fewtabtrue

{\count255=\time\divide\count255 by 60
\xdef\hourmin{\number\count255}
\multiply\count255 by-60\advance\count255 by\time
\xdef\hourmin{\hourmin:\ifnum\count255<10 0\fi\the\count255}}
\def\ps@draft{\let\@mkboth\@gobbletwo
    \def\@oddhead{}
    \def\@oddfoot
       {\hbox to 7 cm{$\scriptstyle Draft\ version:\ \draftdate$
       \hfil}\hskip -7cm\hfil\rm\thepage \hfil}
    \def\@evenhead{}\let\@evenfoot\@oddfoot}

\def\ceqno{\global\@fewtabfalse
    \ifcase\@eqcnt \def\@tempa{& & &}\or \def\@tempa{& &}
      \or \def\@tempa{&}
      \or\def\@tempa{}\fi\@tempa
{\rm(\theequation)}}

\def\aeqno#1{\global\@fewtabfalse
    \ifcase\@eqcnt \def\@tempa{& & &}\or \def\@tempa{& &}
      \or \def\@tempa{&}
      \or\def\@tempa{}\fi\@tempa
{\rm(\theequation,#1)}}

\def\label#1{\ifnum\draftcontrol=1
 \global\def\draftnote{$\scriptstyle #1$}\fi
 \@bsphack\if@filesw {\let\thepage\relax
   \def\protect{\noexpand\noexpand\noexpand}%
\xdef\@gtempa{\write\@auxout{\string
      \newlabel{#1}{{\@currentlabel}{\thepage}}}}}\@gtempa
   \if@nobreak \ifvmode\nobreak\fi\fi\fi
  \@esphack}

\def\alabel#1#2{\label{#1}\global\@fewtabfalse
    \ifcase\@eqcnt \def\@tempa{& & &}\or \def\@tempa{& &}
      \or \def\@tempa{&}
      \or\def\@tempa{}\fi\@tempa
{\hbox to 3cm{\phantom{\rm(\theequation,#2)}
\draftnote \hfil}\hskip -3cm {\rm(\theequation,#2)}}}

\def\clabel#1{\label{#1}\global\@fewtabfalse
    \ifcase\@eqcnt \def\@tempa{& & &}\or \def\@tempa{& &}
      \or \def\@tempa{&}
      \or\def\@tempa{}\fi\@tempa
{\hbox to 3cm{\phantom{\rm(\theequation)}
\draftnote \hfil}\hskip -3cm{\rm(\theequation)}}}

\def\eqnarray{\def\draftnote{{}}\global\@fewtabtrue
\stepcounter{equation}\let\@currentlabel=\theequation
\global\@eqnswtrue
\global\@eqcnt\z@\tabskip\@centering\let\\=\@eqncr
$$\halign to \displaywidth\bgroup\@eqnsel\hskip\@centering\@eqcnt\z@
  $\displaystyle\tabskip\z@{##}$&\global\@eqcnt\@ne
  \hskip 1\arraycolsep \hfil${##}$\hfil
  &\global\@eqcnt\tw@ \hskip 1\arraycolsep
$\displaystyle\tabskip\z@{##}$
\hfil  \tabskip\@centering&\global\@eqcnt\thr@@\llap{##}\tabskip\z@
\cr}

\def\endeqnarray{\@@eqncr\egroup
      \global\advance\c@equation\m@ne$$\global\@ignoretrue}

\def\@eqnnum{\hbox to 3cm{\phantom{\rm(\theequation)} \draftnote
                         \hfil}\hskip -3cm {\rm(\theequation)}}

\def\@@eqncr{\let\@tempa\relax
    \ifcase\@eqcnt \def\@tempa{& & &}\or \def\@tempa{& &}
      \or \def\@tempa{&}
      \or\def\@tempa{}
\fi\@tempa
\if@eqnsw
\if@fewtab\@eqnnum\fi
\stepcounter{equation}\fi\global
\@eqnswtrue\global\@eqcnt\z@\global\@fewtabtrue\cr}

\def\draftcite#1{\ifnum\draftcontrol=1#1\else{}\fi}

\def\@lbibitem[#1]#2{\item{}\hskip -3cm \hbox to 2cm
{\hfil$\scriptstyle\draftcite{#2}$}\hskip
1cm[\@biblabel{#1}]\if@filesw
     {\def\protect##1{\string ##1\space}\immediate
      \write\@auxout{\string\bibcite{#2}{#1}}}\fi\ignorespaces}

\def\@bibitem#1{\item\hskip -3cm \hbox to 2cm
{\hfil $\scriptstyle\draftcite{#1}$}\hskip 1cm
\if@filesw \immediate\write\@auxout
       {\string\bibcite{#1}{\the\value{\@listctr}}}\fi\ignorespaces}

\def\nsection#1{\section{#1}}

\def\nappendix#1{\vskip 1cm\no{\bf Appendix #1}\def\thesection{#1}
\setcounter{equation}{0}}

\font\tendl=msbm10  scaled \magstep1
\font\sevendl=msbm7 scaled \magstep1
\font\fivedl=msbm5 scaled \magstep1
\font\tengl=eufm10  scaled \magstep1
\font\sevengl=eufm7 scaled \magstep1
\font\fivegl=eufm5 scaled \magstep1

\newfam\dlfam  
\textfont\dlfam=\tendl \scriptfont\dlfam=\sevendl
\scriptscriptfont\dlfam=\fivedl
\newfam\glfam  
\textfont\glfam=\tengl \scriptfont\glfam=\sevengl
\scriptscriptfont\glfam=\fivegl

\def\draftdate{\number\month/\number\day/\number\year\ \ \ \hourmin }

\global\def\draftcontrol{0}
\catcode`\@=12
\def\tilde{\widetilde}
\def\hat{\widehat}

\documentstyle[11pt,epsf]{article}

\def\theequation{\arabic{equation}} 

\newcommand{\be}{\begin{eqnarray}}
\newcommand{\en}{\end{eqnarray}\vs 0.5 cm}
\newcommand{\non}{\nonumber}
\newcommand{\no}{\noindent}
\newcommand{\vs}{\vskip}

\newcommand{\NR}{{{\bf R}}}

\newcommand{\NC}{{{\bf C}}}

\newcommand{\NZ}{{{\bf Z}}}

\newcommand{\Nt}{{\bf t}}

\newcommand{\Ng}{{\bf g}}
\newcommand{\qq}{\begin{eqnarray}}

\newcommand{\da}{\partial}
\newcommand{\ee}{{\rm e}}

\newcommand{\qqq}{\end{eqnarray}}

\newcommand{\tr}{\hbox{tr}}

\newcommand{\CA}{{\cal A}}

\newcommand{\CC}{{\cal C}}

\newcommand{\CF}{{\cal F}}

\newcommand{\CH}{{\cal H}}

\newcommand{\CM}{{\cal M}}

\newcommand{\CO}{{\cal O}}
\newcommand{\CP}{{\cal P}}
\newcommand{\CQ}{{\cal Q}}

\newcommand{\CU}{{\cal U}}
\newcommand{\CV}{{\cal V}}
\newcommand{\CW}{{\cal W}}
\newcommand{\CX}{{\cal X}}

\newcommand{\m}{\hspace{0.025cm}}

\newcommand{\hf}{{_1\over^2}}

\begin{document}
\title{Canonical Quantization of the Boundary Wess-Zumino-Witten 
Model}
\author{\ 
\\Krzysztof Gaw\c{e}dzki \\ C.N.R.S., I.H.E.S.,
F-91440  Bures-sur-Yvette, France\\
and Laboratoire de Physique, ENS-Lyon,\\46, All\'ee d'Italie, F-69364 Lyon, 
France  \\ \\Ivan T. Todorov \\
Institute for Nuclear Research
and Nuclear Energy,\\ Tsarigradsko
Chaussee 72, BG-1784, Sofia, Bulgaria\\
\\Pascal Tran-Ngoc-Bich\\ 7, rue Alexis Carrel, F-13004 Marseille, 
France}
\date{ }
\maketitle

\vskip 0.3cm
\vskip 1 cm

\begin{abstract}
\vskip 0.3cm

\noindent  

\end{abstract}

We present an analysis of the canonical 
structure of the Wess-Zumino-Witten theory 
with untwisted conformal boundary conditions. 
The phase space of the boundary theory on a
strip is shown to coincide with the phase 
space of the Chern-Simons theory on a solid 
cylinder (a disc times a line) with two Wilson 
lines. This reveals a new aspect of the relation 
between two-dimensional boundary conformal field 
theories and three-dimensional topological 
theories. A decomposition of the Chern-Simons
phase space on a punctured disc in terms
of the one on a punctured sphere and of coadjoint 
orbits of the loop group easily lends itself 
to quantization. It results in a description 
of the quantum boundary degrees of freedom 
in the WZW model by invariant tensors in a 
triple product of quantum group representations. 
The bulk primary fields of the WZW model are 
shown to combine, in the action on the space
of states of the boundary theory, the usual 
vertex operators of the current algebra with 
monodromy acting on the quantum group invariant 
tensors. We present the details of this 
construction for the spin $1/2$ fields in the 
$SU(2)$ WZW theory, establishing their locality 
and computing their 1-point functions.
\vskip 1.3cm

\nsection{Introduction}
\vskip 0.5cm

Two-dimensional boundary conformal field theory is 
a subject under intense study. Models of the 
theory find multiple applications in the analysis 
of two- or 1+1-dimensional critical phenomena 
in the presence of physical boundaries \cite{Card}, 
localized impurities in a metal \cite{Affl}, or point 
contacts in quantum Hall devices or quantum wires
\cite{FLS}. In string theory they describe branes on 
which open strings end \cite{SR,SFW}. A boundary
conformal field theory model is a quantum field theory
in a half space that exhibits invariance under the 
conformal transformations preserving the boundary. 
In two dimensions such transformations form an infinite 
dimensional group of reparametrizations of a line. 
This rich symmetry (or its generalizations) are 
powerful enough to allow in many cases a classification 
of possible solutions, similarly as in the simpler case
without boundary \cite{BPZ}. Although much progress has been
achieved in understanding boundary CFT's since the seminal 
paper of Cardy \cite{Card}, much more remains 
to be done. The structure involved in the boundary 
CFT's is richer than in the bulk theory and classification 
program involves new notions \cite{PetZub,Zuber}. 
One approach that offered a conceptual insight into 
the properties of correlation functions of boundary 
conformal models consisted of relating them to boundary 
states in three-dimensional topological field theories 
\cite{FFFS1,FFFS2}. In the simplest case of the boundary 
Wess-Zumino-Witten (WZW) models (conformal sigma models 
with a group $G$ as a target \cite{WittWZW}), the topological 
three-dimensional model appears to be the group $G$ 
Chern-Simons (CS) gauge theory \cite{WittCS}. 
\vskip 0.3cm

The purpose of the present paper is to demonstrate 
another facet of the relationship between the boundary 
WZW models and the CS theory, already present at the
classical level. We shall discover it by analyzing 
the structure of the phase space of the WZW model 
with the most symmetric boundary conditions. These, so 
called ``untwisted'', boundary conditions restrict the 
boundary values of the classical fields of the model to 
fixed conjugacy classes in $G$ which are labeled by weights 
of the Lie algebra $\Ng$ of $\,G$. Such boundary conditions 
reduce to the Dirichlet conditions for toroidal targets. 
We shall show that the phase space of the WZW model on 
a strip with the untwisted boundary conditions is isomorphic 
to the phase space of the CS theory on a disc $D$ times 
the time line $\NR$, with two timelike Wilson lines 
corresponding to the weights labeling the boundary conditions.
This generalizes the case with one Wilson line which 
is well known to reproduce the coadjoint orbits 
of the (central extension) of the loop group $LG$
\cite{EMSS}. The isomorphism to the CS theory on $D\times\NR$
is another manifestation of the chiral character of 
the boundary CFT which has half of the bulk symmetries
and correlation functions given by special chiral conformal 
blocks on a double surface. The CS theory (certainely
abelian but possibly nonabelian) describes the long range 
degrees of freedom in the physics of Quantum Hall Effect 
\cite{FK,FPSW,CGT}, with Wilson lines representing excited
Laughlin vortices. Since the disc geometry appears naturally 
in material samples, our identification raises a possibility 
of new applications of boundary CFT to condensed matter physics. 
\vskip 0.3cm

The phase space of the CS theory on $D\times\NR$ with 
two Wilson lines may be decomposed in terms of the phase 
space of the CS theory on $S^2\times\NR$ with three Wilson 
lines and the coadjoint orbits of the loop group,
with one Wilson line indexed by the same weight 
as the loop group orbit. This is the realization 
of the phase space of the boundary WZW model that we 
analyze in detail\footnote{A direct discussion of the
CS theory on $D\times\NR$ with two Wilson lines will be 
presented elsewhere.}. The symplectic structure of the CS 
theory on $\Sigma\times\NR$, where $\Sigma$ is a compact 
surface without boundary, with timelike Wilson lines, has 
been studied in a number of mathematical papers, see 
e.g.\,\,\cite{H,J,JW}. The phase space of 
the theory is composed of flat connections on punctured 
$\Sigma$, modulo gauge transformations, with prescribed 
conjugacy classes of the holonomy around the punctures. 
We shall make use of the paper \cite{AlMal} that contains 
the calculation of the symplectic 
structure of the phase space in terms of the holonomy 
of the flat connection. This presentation of the CS phase 
spaces allows us to identify a factor in the phase space 
of the boundary WZW model as the CS phase space for 
the $S^2\times\NR$ geometry with three timelike Wilson 
lines. The latter space, as was realized in \cite{AlMal},
may be also described in terms of the Poisson-Lie geometry.
It is isomorphic to a reduction of a product of symplectic 
leaves of the Poisson-Lie group $G^*$ dual to $G$ 
equipped with the $r$-matrix Poisson-Lie group structure 
\cite{STS}. One reduces the product of the leaves with 
respect to the diagonal Poisson-Lie ``dressing'' action 
of $G$. 
\vskip 0.3cm

The above identifications permit a decomposition of the phase 
space of the boundary WZW model in terms of the coadjoint 
orbits of $LG$ and of the reduced products of symplectic 
leaves of $\,G^*$. The main point of the above analysis
is that in such a presentation the WZW phase space
may be easily quantized. The coadjoint orbits of the loop 
group give rise upon quantization to the unitary projective 
representations of $LG$ (or of the corresponding affine 
current algebra $\hat{\Ng}$). Geometric quantization of 
the phase-space of the CS theory on $S^2\times\NR$ with 
Wilson lines produces the space of conformal blocks of the WZW 
theory on punctured $S^2$. As for the symplectic leaves of $G^*$,
they may be quantized to irreducible representations of the 
quantum deformation $\CU_q(\Ng)$ of the enveloping algebra 
of $\Ng$. The diagonal reduction of the product 
of symplectic leaves imposes on the quantum level 
a restriction to invariant tensors of the product 
of quantum group representations. It is indeed well known 
that the conformal blocks of the WZW model on a punctured 
sphere may be identified with (``good'') invariant tensors 
of the quantum group \cite{FW}. This is, in fact, the 
way by which the quantum group tensors entered the analysis 
of bulk CFT's. Their appearance in the boundary theory
is even more natural as in the latter they describe
directly a part of the physical degrees of freedom.
\vskip 0.3cm

A concrete realization of the space of quantum states 
of the boundary WZW model in geometric terms would not 
be very useful if it did not lead to a natural description 
of the rest of the quantum field theory structure.
We then show how to use our geometric approach to construct
the action of the bulk primary fields in the Hilbert
space of the boundary model. The bulk operators are built by
combining the vertex operators acting between the unitary
representations of the current algebra $\hat{\Ng}$
with ``monodromy'' expressed as a combination 
of quantum group generators and intertwiners that acts in the 
spaces of invariant quantum group tensors. We make this construction 
explicit for the case of the $SU(2)$ group and spin $1/2$ bulk 
fields using free field realizations of the current algebra 
and of the quantum group representations. The main result here
is the proof of locality of the constructed fields. Our 
analysis does not exhaust the algebraic content of the 
boundary WZW $SU(2)$ model. We do not discuss the higher 
spin bulk operators (they could be constructed along 
similar lines as for the spin $1/2$ fields or by fusing 
the latter). Neither do we discuss the boundary operators, 
although the ones which do not change boundary conditions 
may be easily obtained from the bulk operators by sending
the insertion point to the boundary. An extension of 
the present approach to boundary changing operators would 
require going beyond the strip geometry of the world-sheet 
analyzed here. Other obvious open problems are an
extension of the analysis to twisted boundary conditions
\cite{SFW} and to other groups.
\vskip 0.3cm

The paper is organized as follows. In Sect.\,\,2, we describe 
the canonical structure of the bulk WZW theory studied in
numerous publications, see \cite{Fadd,AlSh,Godd,Chu,FHT,BFP}. 
Our exposition follows closely that of \cite{COQG,FG}.
In particular, we analyze the decomposition of the bulk phase
space into chiral components. In Sect.\,\,3, we describe 
the phase space of the boundary WZW model stressing 
the similarities and differences with the chiral sector
of the bulk theory. In Sect.\,\,4.1, we recall the results
of \cite{AlMal} about the phase space of the CS theory
on $S^2\times\NR$ with three Wilson lines and identify
the latter space with the phase space of the boundary degrees
of freedom in the boundary WZW model. In Sect.\,\,4.2,
we show how to identify the complete phase space of the boundary 
theory with  the phase space of the CS theory on $D\times\NR$ 
with two Wilson lines. Sect.\,\,5 recalls the relations 
between the CS phase space and Poisson-Lie symplectic 
leaves following again the results of \cite{AlMal}.
Sect.\,\,6 discusses quantization of the building
blocks of the boundary theory. In Sect.\,\,6.1, we describe 
the Hilbert space of states in the boundary theory that factors 
into the unitary representations of the current algebra and 
the finite-dimensional spaces of 3-point conformal blocks. 
In Sect.\,\,6.2, we recall the free field realizations 
of the unitary representations of the current algebra 
$\hat{su}(2)$ \cite{Wak} and of the spin $1/2$ vertex 
operators \cite{BerFel}. Sect.\,\,6.3 is devoted to
similar constructions for the $\,\CU_q(su(2))$ quantum group
\cite{FG}. We obtain a ``free field'' realization
of the spaces of invariant quantum group tensors and of
the action on it of the monodromy operators. In Sect.\,\,7, 
we make use of the preceding constructions to describe 
the action of the bulk spin 1/2 fields in the Hilbert space
of the boundary theory. We define the quantum bulk fields 
in Sect.\,\,7.1 and check their locality in Sects.\,\,7.2 
and 7.3. Finally, in Sect.\,\,7.4, we compute some simple 
matrix elements of these operators. Appendices
establish two algebraic identities used in the text.
\vskip 0.9cm

\nsection{Canonical quantization of the bulk WZW model}
\vskip 0.5cm

On the classical level, the WZW model is specified by the
action functional of classical fields. Its symmetry structure 
is identified by examining field transformations mapping
classical solutions (i.e.\,\,extremal points of the action)
to classical solutions. Quantization of the model 
is performed in the way that preserves the classical
symmetries.
\vskip 0.3cm

Let us start by reminding how this is done for the
WZW model in the bulk, see \cite{COQG,FG}. As the two-dimensional 
(Minkowski) space-time $M$ we shall take the cylinder $\,\NR\times 
S^1\,$ with the coordinates $(t,\m x\,{\rm mod}\,2\pi)$. We shall 
also use the light-cone coordinates $x^\pm= x\pm t$ on $M$ in which 
the metric takes the form $ds^2=dx^+dx^-$. The fields of the WZW 
model on $M$ are the maps $\,g:M\rightarrow G\m$, \, see
Fig.\,\,1,

\leavevmode\epsffile[0 -20 320 193]{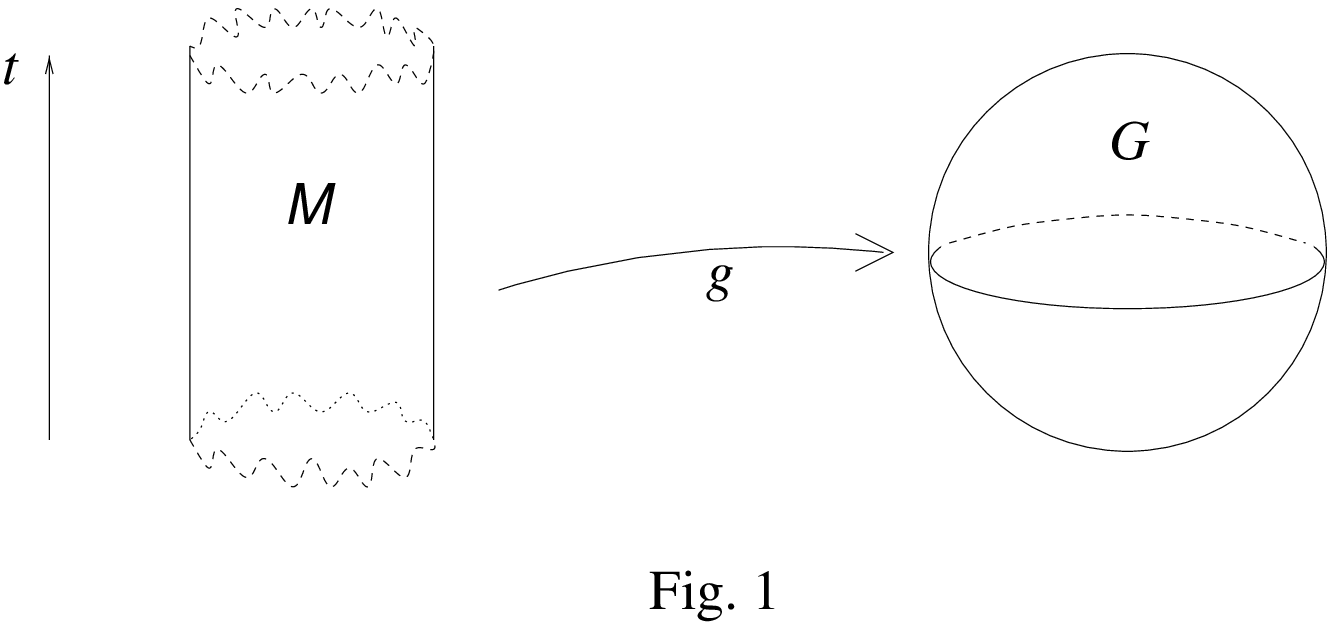}

\noindent where $G$ is a compact group that we shall 
take simple, connected and simply connected. The action 
of the model is given by the expression
\qq
S(g)\ =\ {_k\over^{4\pi}}\int\limits_{M}
[\,\tr\,(g^{-1}\da_{_+}g)(g^{-1}\da_{_-}g)
\,\,dx^+dx^-\ +\ g^*\omega\,]\,,
\label{act}
\qqq
where $k$ is the coupling constant (the "level" of the model),
$\tr$ stands for the Killing form on the
(complexification of the) Lie algebra $\Ng$ of $G$ (normalized 
to give length square 2 of the long roots), $\omega$ 
is a 2-form on $G$ satisfying
\qq
d\omega(g)\ =\ {_1\over^3}\,\,\tr\,(g^{-1}dg)^3
\,\equiv\,\theta(g)\,
\label{beta}
\qqq
and $g^*\omega$ denotes the pull-back of $\omega$ (we use the symbol 
$g$ to denote both the field mapping $M$ to $G$ and an element 
of the group $G$). The last equality requires 
a comment, since the 3-form $\theta$ on $G$ is closed but not 
exact so that no 2-form $\omega$ satisfying
Eq.\,\,(\ref{beta}) exists globally on $G$. To simplify the more 
complex story (see e.g.\,\,Sect.\,\,7 of \cite{Istam}), 
we shall assume that the values of the field $g$ belong to an open 
subset of $\m G$ on which one can define such a 2-form. 
On the space-time without boundary, local variations of
the action and, consequently, also the classical equations,
will be independent of the choice of $\omega$. 
\vskip 0.3cm

It will be convenient to rewrite the action (\ref{act}) 
in the first order formalism, see \cite{COQG}. To this end, 
we introduce additional Lie-algebra valued coordinates
$\xi_{_\pm}$ (which will represent the values of 
field derivatives) and we define a 2-form $\alpha$ 
on the extended space $\,P\,\equiv\,M\times G\times\Ng^2\m$,
\qq
\alpha\ =\ {_k\over^{4\pi}}\,[\,\tr\,\xi_{_+}\xi_{_-}\,\,
dx^+dx^-\ -\ i\,\m\tr\,\xi_{_+}(g^{-1}dg)\,dx^+\ 
+\ i\,\m\tr\,\xi_{_-}(g^{-1}dg)\, dx^-\,
\ +\ \omega(g)\,]\,.
\label{alpha}
\qqq
The first order action takes the simple form of the space-time 
integral of a pull-back of the form $\alpha\m$:
\qq
S(\Phi)\ =\ \int\limits_{M}\Phi^*\alpha\,,
\label{1act}
\qqq
where $\,\Phi=(I,\m g,\m \xi_{_+},\xi_{_-})$ maps the space-time 
to $\m P$ \,($I\m$ stands here for the identity map of $M$). 
If the new fields $\xi_{_\pm}$ 
are given by the light-cone derivatives of $g$,
\qq
\xi_{_\pm}\ =\ {_1\over^i}\,g^{-1}\da_{_\pm}g\,,
\label{foe}
\qqq 
then the first order action (\ref{1act}) reduces to the original
expression (\ref{act}). The first order formalism is, however, 
more geometric. For example, the variation of the action
(\ref{1act}) takes the form 
\qq
\delta S(\Phi)\ =\ \int\limits_M\Phi^*(\iota_{_{\delta\Phi}}
\,d\alpha)\,,
\label{fov}
\qqq
where $\iota_{_{\delta\Phi}}$ denotes the interior product 
(contraction) with the vector field $\delta\Phi$
giving the infinitesimal variation of $\Phi$ ($\delta\Phi$
is defined on the range of $\Phi$ and is tangent to $P$).  
Consequently, the classical equations in the first order 
formalism take the form
\qq
\Phi^*(\iota_{_X}\,d\alpha)\ =\ 0\quad\ {\rm for\ every\ vector\ 
field}\ X\ {\rm on}\ P\m. 
\label{cfoe}
\qqq
These equations are equivalent
to the relations (\ref{foe}) supplemented with the variational equation
$\delta S(g)=0$ for the second-order action. The latter  
requires that
\qq
\da_{_-}(g\,\da_{_+}g^{-1})\ =\ 0\,,
\label{ceq}
\qqq
or, equivalently, that \,$\da_{_+}(g^{-1}\da_{_-}g)=0$.
\vskip 0.3cm

Eq.\,\,(\ref{ceq}) is easy to solve. The solutions on 
the cylinder $M$ decompose into the left- and right-moving 
components (generalizing the decomposition of the solutions
of the linear wave equation):
\qq
g(t,x)\ =\ g_{_L}(x^+)\,\, g_{_{R}}(x^-)^{-1}\,,
\label{dem}
\qqq
where the chiral fields $\,g_{_{L,R}}$ are arbitrary $G$-valued 
maps on the real line satisfying 
\qq
g_{_{L,R}}(y+2\pi)\ =\ g_{_{L,R}}(y)\,\gamma
\qqq
with the same monodromy $\gamma\in G$. \,By the right multiplication
of $g_{_L}$ and $g_{_R}$ by the same element of $G$, a change
that does not effect the solution, one may reduce the monodromy 
$\gamma$ to the Cartan subgroup $T\subset G$ or, even more,
to the form $\,\gamma=\ee^{\m2\pi i\m\tau}\m$, \,where $\m\tau\m$
belongs to the positive Weyl alcove $\m\CA_{_W}$ in the Cartan
algebra $\Nt$. Nevertheless, it will be sometimes convenient to work 
with general $\gamma$. 
\vskip 0.3cm

The space $\,\CP\,$ of the classical solutions given explicitly 
by Eq.\,\,(\ref{dem}) forms the phase space of the WZW model on 
the cylinder. As usual, the phase space comes with the canonical 
symplectic structure. The symplectic form $\Omega$ on $\CP$ may 
be conveniently expressed in the first order formalism, 
see e.g.\,\,\cite{COQG}. Namely,
\qq
\Omega(\delta_1\Phi,\,\delta_2\Phi)\ =\ 
\int\limits_{M_t}\Phi^*(\iota_{_{\delta_2\Phi}}
\,\iota_{_{\delta_1\Phi}}\,d\alpha)\,,
\label{Om}
\qqq
where $\,M_t$ denotes the constant time section of $M$.
The integral on the right hand side is $t$-independent since 
the integrated form is closed. Explicitly \cite{COQG}:
\qq
\Omega\ =\ {_k\over^{4\pi}}\int_{_0}^{^{2\pi}}
\hspace{-0.2cm}\tr\,\,[\m-\delta(g^{-1}
\da_tg)\,g^{-1}\delta g\,+\,2\,(g^{-1}\da_{_+}g)\,(g^{-1}
\delta g)^2\m]\,\,dx\,,
\label{Omega}
\qqq
where $\delta$ denotes here the exterior derivative on $\CP$
and the $x$-integral is performed with fixed $t$. 
\vskip 0.3cm

Although we have originally assumed that the group $G$ was 
compact, in all the formulae above, we could replace $G$ by its 
complexification.  The phase space $\CP$ would then become 
a complex symplectic manifold. Below, we shall work in 
the complex context whenever more convenient.
\vskip 0.3cm

The symplectic structure of $\CP$ allows to assign to functions
$\CF$ on $\CP$ the Hamiltonian vector fields $\CX_{_{\CF}}$ such that 
$\m d\CF=\iota_{_{\CX_{_\CF}}}\Omega\m$ and to define the Poisson bracket 
$\,\{\CF,\CF'\}=\CX_{_\CF}(\CF')\m$ of functions on the phase space. 
Some equal-time 
Poisson brackets are easy to compute. For example, if $g(t,x)_{_1}$ 
and $g(t,x)_{_2}$ denote the matrices $\m g(t,x)\otimes I\m$ 
and $\m I\otimes g(t,x)\m$ in a fixed representation of $\m G$, 
with a similar notation for the Lie-algebra valued fields, then
\qq
&&\hbox to 6cm{$\Big\{\,g(t,x)_{_1}\,\m,\,\,g(t,x')_{_2}\m\Big\}$
\hfill}=\ 0\,,\cr\cr
&&\hbox to 6cm{$\Big\{\,g(t,x)_{_1}\,\m,\,\,(g^{-1}\da_t g)(t,x')_{_2}
\m\Big\}$\hfill} =\ -\m{_{4\pi}\over^k}\,\delta(x-x')\,\,g(t,x)_{_1}
\,\,C_{_{12}}\,,\cr\cr
&&\hbox to 6cm{$\Big\{\,(g^{-1}\da_tg)(t,x)_{_1}\,\m,\,\,(g^{-1}\da_t 
g)(t,x')_{_2}\m\Big\}$\hfill}=\ {_{8\pi}\over^k}\,\delta(x-x')\,\,[\,
C_{_{12}}\,,\m\,(g^{-1}\da_{_+}g)(t,x)_{_1}\m]\,,
\label{ccr}
\qqq
where the matrix product is implied on the left hand side and
$\,C_{_{12}}\,$ stands for the matrix representing the Casimir 
element $\sum t^a\otimes t^a\m\in\m\Ng\otimes\Ng$, with the generators 
$t^a$ of the Lie algebra $\Ng$ such that $\,\tr\, t^a t^b=
\delta^{ab}$.
\vskip 0.3cm

It is easy to identify the symmetry structure of the WZW 
theory on the cylinder. First, the loop group $LG$ composed 
of the periodic maps $h$ from the line to $G$ with period $2\pi$ 
act on the phase space $\CP$ in two ways by
\qq
\quad g(x^+,x^-)\ \,\longmapsto\,\ h(x^+)\, g(x^+,x^-)\,,\qquad\qquad
g(x^+,x^-)\ \,\longmapsto\,\ g(x^+,x^-)\, h(x^-)^{-1}\,\,\qquad
\qqq
preserving the symplectic structure. On the infinitesimal level, 
these actions are generated by the currents\footnote{More
precisely, the functions $\,\CF=\pm\m{_1\over^{2\pi}}\int_{_0}^{^{2\pi}}
\hspace{-0.15cm}\tr\,\,\delta\Lambda(x^\pm)\,J_{_{L,R}}(t,x)\,
dx\,$ generate the Hamiltonian vector fields corresponding
to the action of the loop group elements $\,h(x^\pm)
=\ee^{-i\m\delta\Lambda(x^\pm)}\m$.}
\qq
J_{_L}\ =\ {{i\m k}}\,g\m\da_+g^{-1}\ =\ 
{{i\m k}}\,g_{_L}\m\da_+g_{_L}^{-1}\,,\qquad
-\m J_{_R}\ =\ -\m{{i\m k}}\, g^{-1}\da_{_-}g\ =\ 
-\m{{i\m k}}\,g_{_R}\m\da_-g_{_R}^{-1}\,\m\quad
\label{curr}
\qqq
which are periodic functions with period $2\pi$ of $x^+$ and $x^-$, 
respectively. Second, there are two commuting actions on $\CP$ 
of the group $Diff_{_+}(S^1)$ of the orientation-preserving 
diffeomorphisms $D$ of the circle $S^1\cong\NR/2\pi\NZ$ given by:
\qq
\quad g(x^+,x^-)\ \,\longmapsto\,\ g(D^{-1}(x^+),\,x^-)\,,\qquad\quad
g(x^+,x^-)\ \,\longmapsto\,\ g(x^+,\m D^{-1}(x^-))\,.\qquad\ \ \,
\qqq
They also preserve the symplectic structure. Their infinitesimal 
versions are generated by the non-vanishing components 
of the energy-momentum tensor\footnote{More precisely, 
$\,\CF=\pm\m{_1\over^{2\pi}}\int_{_0}^{^{2\pi}}\hspace{-0.15cm}
\delta\xi(x^\pm)\,T_{_{L,R}}(t,x)\,dx\,$ generate the 
Hamiltonian vector fields corresponding to the action
of the diffeomorphisms $\,D=\ee^{\m\delta\xi(x^\pm)\,\da_\pm}\m$.} 
\qq
T_{_L}\,=\,-\m{_k\over^{2}}\,\tr\,(g\,\da_{_+}g^{-1})^2\,
=\,{_1\over^{2k}}\,\tr\,\, J_{_L}^2\,,\qquad
-\m T_{_R}\,=\,{_k\over^{2}}\,\tr\,(g^{-1}\da_{_-}g)^2\,
=\,-\m{_1\over^{2k}}\,\tr\,\, J_{_R}^2\,.\ \ \ 
\label{emt}
\qqq
These are the infinite-dimensional symmetries of the theory.
\vskip 0.3cm

In order to achieve a formulation of the classical WZW model
that lends itself more easily to quantization, it is convenient 
to express the symplectic structure of the phase space $\CP$ 
in terms of the chiral components $g_{_{L,R}}$ of the classical 
solutions. One obtains:
\qq
\Omega\ =\ \Omega_{_L}\,-\ \Omega_{_R}
\label{OmLR}
\qqq
\vskip -0.4cm
\noindent where 
\qq
\Omega_{_L}\,=\,{_k\over^{4\pi}}\Big[\int_{_0}^{^{2\pi}}
\hspace{-0.2cm}\tr\,\,(g_{_L}^{-1}\delta g_{_L})\,\da_x(g_{_L}^{-1}
\delta g_{_L})\m\, dx\ \,+\,\ \tr\,\,(g_{_L}^{-1}\delta g_{_L})(0)
\,(\delta\gamma)\,\gamma^{-1}\Big]
\label{OmL}
\qqq
and $\m\Omega_{_R}$ is given by the same formula with $g_{_R}$
replacing $g_{_L}$. The reversed sign in front of $\,\Omega_{_R}$
is the source of the negative signs in front of $J_{_R}$ and
of $\,T_{_R}$ above. The chiral 2-forms $\m\Omega_{_{L,R}}$ 
on $\CP$ are not closed. An easy computation gives:
\qq
\delta\Omega_{_L}\ =\ {_k\over^{4\pi}}\,\theta(\gamma)\,,
\qqq
where, as before, $\theta(\gamma)={_1\over ^3}\,\tr\,(\gamma^{-1}d
\gamma)^3$. If we restrict, however, the monodromy of the 
twisted-periodic fields $g_{_L}$ to be of the form $\gamma
=\ee^{\m2\pi i\m\tau}$ with $\tau\in\CA_{_W}$ then the forms 
$\Omega_{_{L,R}}$ become closed and define the symplectic 
structure on the chiral components $\CP_{_{L,R}}$ of the phase 
space composed of fields $g_{_L}$ and $g_{_R}$ with the restricted 
monodromy. 
\vskip 0.3cm

One may also proceed differently \cite{COQG} keeping 
the monodromies general and introducing modified forms
\qq
\tilde\Omega_{_L}\ =\ \Omega_{_L}
\,-\,\rho(\gamma)\,,\qquad\tilde\Omega_{_R}\ 
=\ \Omega_{_R}\,-\,\rho
(\gamma)\,,
\label{modf}
\qqq
where $\m\rho\m$ is a 2-form on $G$. The decomposition
$\m\Omega=\tilde\Omega_{_L}-\tilde\Omega_{_R}$ still holds 
since the $\rho$-terms cancel. If the form $\rho$
were such that $d\rho=\theta$ then the modified 2-forms 
$\tilde\Omega_{_{L,R}}$ would be closed.
Note that we recover for $\rho$ the same condition as for the 2-form
$\omega$ entering the action of the model, see Eq.\,\,(\ref{beta}). 
Of course, as before, that condition cannot be satisfied globally.
In the complex setup, a convenient solution is to consider
only generic monodromies that may be parametrized by the Gauss 
decomposition $\gamma=\gamma_{_-} \gamma_{_+}^{-1}$ with $\gamma_{_\pm}$ 
in the Borel subgroups $B_{_\pm}=N_{_\pm} T\subset G$. $\,N_{_\pm}$ 
denote the nilpotent subgroups of $B_{_\pm}$ and $\,\gamma_{_+}$ and 
$\gamma_{_-}^{-1}$ are taken with coinciding components 
in the Cartan subgroup $T$. \,The choice
\qq
\rho(\gamma)\ =\ \tr\,\m(\gamma_-^{-1}d\gamma_-)\m
(\gamma_+^{-1}d\gamma_+)\,,
\label{omg}
\qqq
fulfills the condition $d\rho=\theta$ rendering the forms 
$\tilde\Omega_{_{L,R}}$ closed and providing symplectic structures
on the spaces $\tilde\CP_{_{L,R}}$ of chiral fields (with the monodromies 
parametrized by the Gauss decomposition). 
\vskip 0.3cm

In \cite{FG} a further change of variables, a classical version
of the so called vertex-IRF (interaction round the face) transformation, 
was described. It decomposed a chiral field into the product of a closed 
loop in $G$, a multi-valued field in the Cartan subgroup and a constant 
element in $G\m$:
\qq
g_{_L}(x)\ =\ h(x)\,\,\ee^{\m i\m\tau\m x}\,\m g_{_0}^{-1}\ \equiv\ 
h_{_L}(x)\,\m g_{_0}^{-1}\,,
\label{dec0}
\qqq
where $h\in LG$, $\,\tau$ belongs to the positive Weyl 
alcove $\m\CA_{_W}\subset\Nt$ (in the complex setup, $\CA_{_W}$ 
should admit arbitrary imaginary parts of $\tau$) and $g_{_0}\in G$. 
For the monodromy of $g_{_L}$, we obtain
\qq
\gamma\ =\ g_{_0}\,\m\ee^{\m2\pi i\m\tau}\, g_{_0}^{-1}\,.
\label{mondr}
\qqq
The parametrization (\ref{dec0}) induces the following 
decomposition of the form $\tilde\Omega_{_L}$:
\qq
\tilde\Omega_{_L}&=&
{_k\over^{4\pi}}\int_{_0}^{^{2\pi}}\hspace{-0.2cm}\tr\,\,
[\m(h^{-1}{\delta}h)\,\da_x(h^{-1}{\delta}h)\,+\, 
2i\m\tau\,(h^{-1}{\delta}h)^2\,-\,2i\,({\delta}\tau)(h^{-1}
{\delta}h)\m]\m\, dx\cr\cr
&&+{_k\over^{4\pi}}\,\tr\,(g_{_0}^{-1}{\delta}g_{_0})\,\ee^{\m2\pi 
i\m\tau}\,(g_{_0}^{-1}{\delta}g_{_0})\,\ee^{-2\pi i\m\tau}\,
+\, k\m i\,\tr\,({\delta}\tau)(g_{_0}^{-1}{\delta}g_{_0})\,
-\,{_k\over^{4\pi}}\,\rho(g_{_0}\,\m
\ee^{\m2\pi i\m\tau}\,g_{_0}^{-1})\cr\cr
&\equiv&\Omega^{LG}\ +\ \Omega^{PL}\,.
\label{deco}
\qqq
The 2-form $\Omega^{LG}$ coincides with the restriction
of the chiral 2-form $\Omega_{_L}$ to the subspace $\m\CP_{_L}$
of fields $h_{_L}$ with monodromy $\gamma=\ee^{\m2\pi i\m\tau}$.
The symplectic space $\m\CP_{_L}$ may be identified with the so called 
``model space'' $\,\CM_{_{LG}}=LG\times\CA_{_W}\m$ of the loop group, 
a symplectic space roughly speaking containing once each coadjoint 
orbit $\CO_{_{LG}}(\tau)$ passing through $\tau$ of the 
(central extension $\hat{LG}$ of the) loop group. More precisely, 
for fixed $\tau$, $\m\Omega^{LG}$ gives the pull-back to $LG$ of the 
Kirillov symplectic form $\Omega^{LG}_\tau$ on $\CO_{_{LG}}(\tau)$. 
We infer that 
\qq
\CP_{_L}\ \cong\ \CM_{_{LG}}
\label{111}
\qqq
as symplectic manifolds. Note the symplectic actions
of the loop group $LG$ and of the Cartan subgroup $T$ on $\CM_{_{LG}}$ 
given, respectively, by $\,(h,\tau)\mapsto(h'h,\tau)\,$ and
$\,(h,\tau)\mapsto(h\m t^{-1},\tau)$. 
\vskip 0.3cm

One may introduce the Darboux coordinates on the (complex version of) 
the model space $\,\CM_{_{LG}}$ using the Gauss decomposition 
of the fields $h_{_L}$. For the $SU(2)$ group the decomposition is 
\qq
h_{_L}\ =\ \left(\matrix{1&{\beta(x)}\cr0&1}\right)\,
\left(\matrix{1&0\cr{w(x)}&1}\right)\,
\left(\matrix{{\psi(x)}&0\cr0&{\psi(x)^{-1}}}\right)\,,
\qqq
see Sect.\,\,5 of \cite{FG}. Defining the modes
\qq
&&\beta(x)\,\equiv\sum\limits_n\beta_n\,\ee^{-inx}\,,\qquad
\gamma(x)\,\equiv\,\sum\limits_n\gamma_n\,\ee^{-inx}
\ =\ ik\,\psi^{-2}(x)\,\partial(\psi^2(x)w(x))\,,\cr
&&\phi(x)\,\equiv\,\phi_{_0}+a_{_0}x+i\sum\limits_{n\not=0}
{_1\over^n}\,a_n\,\ee^{-inx}\ =\ 2i\m\zeta\,\ln{\psi(x)}
\qqq
for $\zeta=\sqrt{_k\over^2}$, one obtains the canonical Poisson brackets
\qq
\{a_n\,,\,a_m\}\ =\ -i\m n\,\delta_{n,-m}\,,
\qquad
\{\phi_{_0}\,,\,a_{_0}\}\ =\ 1\,,
\qquad\{\beta_n\,,\,\gamma_n\}\ =\ -i\,\delta_{n,-m}
\label{ccpb}
\qqq
with the other brackets vanishing. In terms of 
$\phi,\,\beta$ and $\gamma$,
\qq
&&J_{_L}\ =\,\left(\matrix{-\zeta\,\partial\phi-
\beta\m\gamma&-ik\,\partial\beta+2\zeta\,\beta\m
\partial\phi+\beta^2\gamma\cr-\gamma&\zeta
\,\partial\phi+\beta\m\gamma}\right),
\label{ffr1}\\ \cr
&&h_{_L}\ =\,\left(\matrix{(\Pi-\Pi^{-1})\psi+\beta\,\psi^{-1}Q
&\beta\,\psi^{-1}\cr\psi^{-1}Q&\psi^{-1}}\right)
\left(\matrix{(\Pi-\Pi^{-1})^{-1}&0\cr0&1}\right),
\label{ffr2}
\qqq
where  $\,\Pi=\ee^{-\pi i\,\zeta^{-1}a_{_0}}$ 
is the monodromy of $\psi(x)$ and
\qq
Q(x)\ =\ {_1\over^{ik\,\Pi}}\hspace{-0.15cm}\int\limits_x^{x+2\pi}
\gamma(y)\m\psi(y)^2\,dy
\label{scch}
\qqq
is the ``screening charge'' appearing when solving for $w$ in terms
of $\gamma$: \,$w=\psi^{-2}Q/(\Pi-\Pi^{-1})$.
\vskip 0.3cm

Whereas the 2-form $\Omega^{LG}$ involves the loop group geometry,
the 2-form $\Omega^{PL}$, with $\rho$ \m given by Eq.\,\,(\ref{omg}), 
is related to the Poisson-Lie geometry. More precisely,
$\Omega^{PL}$ defines a symplectic form on the space 
$\,\CM^{PL}_{_G} =G\times\CA_{_W}$ 
of pairs\footnote{More precisely, one should consider 
the space of quadruples $(g_{_0},\tau,\gamma_\pm)$ s.\,\,t. $g_{_0}\m
\ee^{\m2\pi i\m\tau}g_{_0}^{-1}=\gamma_{_-}\gamma_{_+}^{-1}$.}
$(g_{_0},\tau)$. For fixed $\tau$, it determines a symplectic 
form $\m\Omega^{PL}_\tau$ on the conjugacy class 
$\m\CC_{_{\tau}}\subset G$ composed of the elements 
$\m g_{_0}\,\ee^{\m2\pi i\m\tau}\m g_{_0}^{-1}\in G$. \,The conjugacy 
classes with the symplectic form $\m\Omega^{PL}_\tau$ may 
be identified with the symplectic leaves of the Poisson-Lie 
group $\,G^*=\{(\gamma_+,\gamma_-)\in B_+
\times B_-\}\,$ dual to the group $G$ equipped with 
the Poisson-Lie structure induced by the standard 
$r$-matrix in $\Ng\otimes\Ng$ \cite{STS}, see \cite{FG}
for a short account. The identification is done via the
map $\m(\gamma_+,\gamma_-)\mapsto\gamma_-\gamma_+^{-1}$.
The conjugacy classes are, of course, the orbits of the 
adjoint action of $G$. More generally, there is a Poisson-Lie
action of $G$ on $\CM^{PL}$ defined by $\,(g_{_0},\tau)\mapsto
(gg_{_0},\tau)\m$. Note also a symplectic action of the Cartan subgroup 
on $\m\CM^{PL}_{_G}$ given by $\,(g_{_0},\tau)\mapsto(g_{_0}
t^{-1},\tau)\m$. 
\vskip 0.3cm

The symplectic leaves of $\m G^*$ play in the Poisson-Lie category 
a role similar to that of the coadjoint orbits 
in the Lie category. The space 
$\m\CM^{PL}_{_G}$ with the symplectic form $\m\Omega^{PL}$ 
may be interpreted as the model space of the Poisson-Lie group $G$, 
containing once each symplectic leaf of $G^*$. The choice (\ref{omg}) 
and the vertex-IRF parametrization (\ref{dec0}) unravel this way 
a hidden Poisson-Lie symmetry of the chiral components of the WZW 
theory \cite{Fadd,AlSh,COQG,BFP}. In particular, we may express the chiral 
phase space $\m\tilde\CP_{_{L}}$ as the symplectic reduction (denoted
by $/\hspace{-0.07cm}/$) of the product of the loop group and 
the Poisson-Lie model spaces by the diagonal action of the Cartan 
subgroup $T$. The reduction imposes the constraint equating the
$\tau$ components in both spaces and takes the orbit space of $T$:
\qq
\tilde\CP_{_L}\ \cong\ (\CM_{_{LG}}\times\CM_{_G}^{PL})\,\Big/
\hspace{-0.15cm}\Big/\, T\ =\ (\CM_{_{LG}}\times_{_{\CA_{_W}}}
\hspace{-0.08cm}\CM_{_G}^{PL})\,\Big/\, T\,.
\label{strL}
\qqq
The representations (\ref{111}) and (\ref{strL}) lend themselves 
easily to the (geometric) quantization.
\vskip 0.3cm

First, the coadjoint orbits of $\hat{LG}$ which pass through 
$\tau=\lambda/k\in\CA_{_W}$, where $k$ is a positive integer 
and $\lambda$ is a weight, may be quantized by the Kirillov-Kostant 
method \cite{Kiril} (for fixed $k$, there is a finite
number of such orbits). Upon quantization, they give rise 
to the irreducible highest weight representations of level 
$k$ of the Kac-Moody algebra $\hat{\bf g}$ ($\simeq$ the Lie 
algebra of $\hat{LG}$) which act in the (infinite-dimensional) 
vector spaces $\CV_{_{k,\lambda}}$. Quantization of the chiral 
phase space $\,\CP_{_{L}}\cong\CM_{_{LG}}$ composed of the chiral 
fields $g_{_L}$ with monodromies of the form $\ee^{\m2\pi i\m\tau}$
is then straightforward and gives the space of quantum states
\qq
\CH_{_L}\ =\ \mathop{\oplus}\limits_{\lambda}\,\CV_{_{k,\lambda}}\,.
\label{chss}
\qqq 
The Kac-Moody algebra action in the representation spaces 
and the Virasoro algebra one, induced from the latter by the 
Sugawara construction, quantize the infinitesimal versions 
of the classical $LG$ and $Diff_{_+}(S^1)$ symmetries of the chiral 
phase space. The space of states of the complete (left-right) 
quantum WZW theory is 
\qq
\CH\ =\ \mathop{\oplus}\limits_\lambda\,\,\CV_{_{k,\lambda}}\otimes\m
\overline{\CV_{_{k,\lambda}}}\,.
\qqq
This mimics the diagonal way in which the classical phase space 
$\CP$ is built from the coadjoint orbits of $\hat{LG}$ in $\CP_{_L}$ 
and in $\CP_{_R}$. The overbar stands for the complex conjugation 
taking into account the opposite symplectic structure 
of the right-handed component of the phase space.
\vskip 0.3cm

In a similar way, the symplectic leaves of $\m G^*$ isomorphic
to the conjugacy classes $\m\CC_{_{\tau}}$ with $\tau=\lambda/k$ 
may be quantized to the irreducible highest-weight 
representations of the quantum deformation $\CU_q(\Ng)$ of the
enveloping algebra of $\Ng$ with the deformation 
parameter $q=\ee^{\m\pi i/(k+h^\vee)}$ ($h^\vee$ denotes 
the dual Coxeter number of the Lie algebra $\m\Ng$). They act 
in the finite-dimensional spaces $\CV_{_{q,\lambda}}$. Quantization 
of the extended chiral phase space $\tilde\CP_{_L}$ gives then 
rise to the space of states
\qq
\tilde\CH_{_L}\ =\ \mathop{\oplus}\limits_\lambda\,
\,\CV_{_{k,\lambda}}\otimes\m \CV_{_{q,\lambda}}\,,
\qqq
which is  the quantum counterpart of the classical decomposition
(\ref{strL}). As explained in \cite{FG} for $G=SU(2)$, one may 
quantize the chiral fields $g_{_L}(x)$ (with general 
monodromies) so that, in the decomposition (\ref{dec0}), $h(x)$ 
becomes a matrix of the ``vertex operators'' of the Kac-Moody 
algebra \cite {TsK} and $g_{_0}^{-1}$ becomes its quantum-group 
counterpart. It should be stressed that in the theory without
boundary, the quantum group degrees of freedom are superfluous 
and serve only to elucidate the chiral structure of the model.
Below, we shall recover a similar coupling of the loop group 
and the quantum group degrees of freedom in the boundary WZW theory. 
In that case, however, both the loop group and the quantum group 
will describe physical degrees of freedom. 
\vskip 0.9cm

\nsection{Phase space of the boundary WZW model}
\vskip 0.5cm

Let us consider now the WZW theory on the space-time $M$ 
in the form of the strip $\NR\times[0,\pi]$, see Fig.\,\,2.

\leavevmode\epsffile[0 -20 320 193]{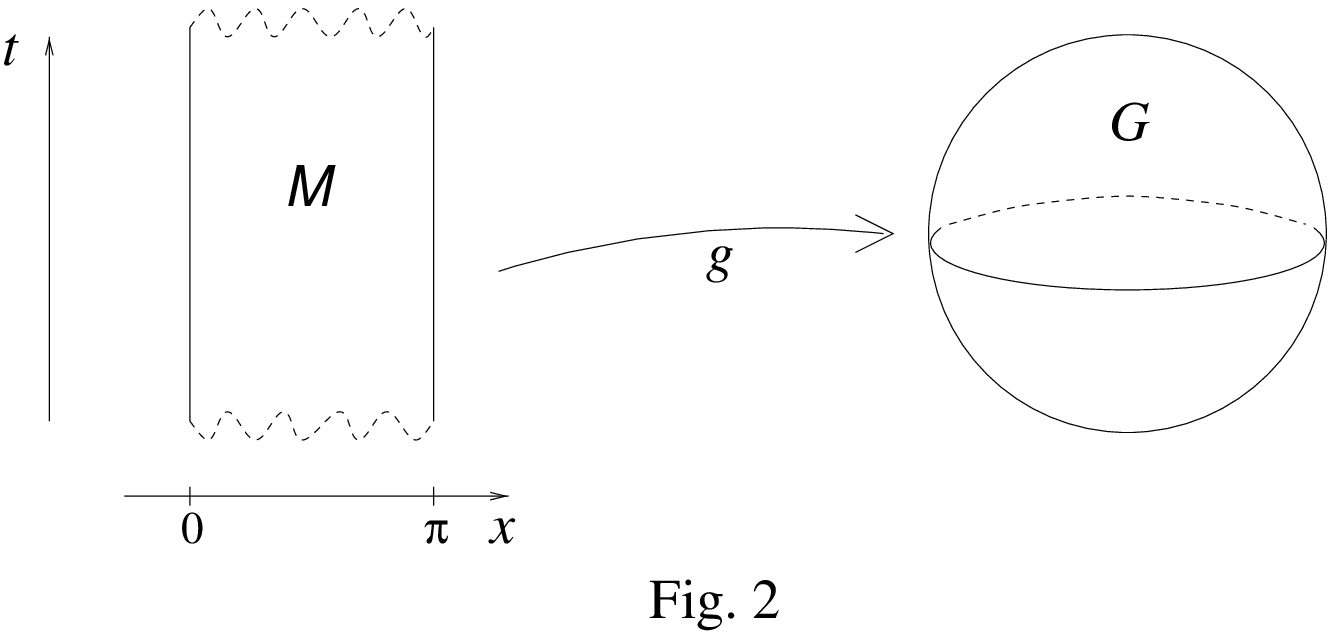}

\noindent Following \cite{Schom}, we shall impose on the fields 
$\,g:M\rightarrow G\,$ the boundary conditions requiring them 
to belong to fixed conjugacy classes on the components of the boundary: 
\qq
g(t,0)\ \in\ \CC_{_{\mu_{_0}}}\,,\qquad g(t,\pi)\ \in\ 
\CC_{_{\mu_{_\pi}}}\,.
\label{bc}
\qqq
As before, $\m\CC_{_{\mu}}=\{\,g_{_0}\,\ee^{\m2\pi i\m\mu}
\m g_{_0}^{-1}\,|\,g_{_0}\in G\}$ and we shall take $\mu$ 
in $\m\CA_{_W}\subset\Nt\m$. This labels the conjugacy classes 
in a one-to-one way. The boundary conditions (\ref{bc}) generalize 
the Dirichlet conditions used for the abelian groups $G$. 
They will permit to preserve in the case with boundary
the infinite dimensional $LG$ and $Diff_{_+}S^1$ symmetries
(there are other choices of boundary conditions with the 
same effect). 
\vskip 0.3cm

The action of the model will be again given by Eq.\,\,(\ref{act}) 
or, in the first order formalism, by Eq.\,\,(\ref{1act}).
Now, however, a boundary term appears in the variation 
of the action:
\qq
\delta S(\Phi)\ =\ \int\limits_M\Phi^*(\iota_{_{\delta\Phi}}
\,d\alpha)\ +\ \int\limits_{\da M}\Phi^*(\iota_{_{\delta\Phi}}\,
\alpha)\,.
\qqq 
The equation $\,\delta S(\Phi)=0\,$ implies then,
besides the bulk relations (\ref{cfoe}), also the boundary
ones which require that
\qq
\Phi^*(\iota_{_X}\,\alpha)\ =\ 0\qquad{\rm along}\qquad\da M 
\qqq
for the vector fields $X$ on $P$ tangent to the boundary
condition surface 
\qq
\{(t,0\}\times\CC_{_{\mu_{_0}}}\times\Ng^2\ \cup\ \{(t,\pi)\}
\times\CC_{_{\mu_{_\pi}}}\times\Ng^2\ \ \subset\ \ P\,\m.
\non
\qqq
Again, the first order variational equations are equivalent
to Eqs.\,\,(\ref{foe}) supplemented by the relations 
$\delta S(g)=0$ for the second order action. The latter, 
besides the bulk equation (\ref{ceq}), require that 
\qq
\tr\,(g^{-1}\delta g)\m(g^{-1}\da_{_+} g)\,dx^+\ -\ 
\tr\,(g^{-1}\delta g)\m(g^{-1}\da_{_-} g)\,dx^-\ =\   
g^*(\iota_{_{\delta g}}\m\omega)
\label{beq}
\qqq
on the vectors tangent to $\da M$. Note that now the choice 
of the form $\omega$ enters the classical equations. 
\vskip 0.3cm

Let us choose the 2-form $\omega$ so that its restrictions 
$\omega_{_\mu}$ to the boundary conjugacy
classes $\CC_{_\mu}$, \m where $\mu=\mu_{_0}$ 
or $\mu=\mu_{_\pi}$, \m take the form
\qq
\omega_{_\mu}(\gamma)\ =\ \tr\,\,(h_{_0}^{-1}dh_{_0})\,
\ee^{\m2\pi i\m\mu}\,(h_{_0}^{-1}dh_{_0})\,\ee^{-2\pi i\m\mu}
\label{omegl}
\qqq
in the parametrization $\,\gamma=h_{_0}\,\ee^{\m2\pi i\m\mu}
\,h_{_0}^{-1}\m$ of $\,\CC_{_\mu}$. Equivalently, 
\qq
\omega_{_\mu}(\gamma)\ =\ \tr\,\,(\gamma^{-1}d\gamma)\,
(1-Ad_\gamma)^{-1}(\gamma^{-1}d\gamma)\ =\ 
\hf\,\tr\,\,(\gamma^{-1}d\gamma)\,{_{1+Ad_g}\over^{1-Ad_g}}\,
(\gamma^{-1}d\gamma)
\qqq
(the linear map \m$(1-Ad_\gamma)$ may be inverted on 
$\gamma^{-1}\delta\gamma$ if $\delta\gamma$ is tangent 
to the conjugacy class of $\gamma$). It is easy 
to check that $d\omega_{_\mu}$ coincides with the restriction 
of the 3-form $\theta$ to $\m\CC_{_\mu}$ so that the choice 
(\ref{omegl}) is consistent with the relation (\ref{beta}). 
For such a choice, the boundary equations (\ref{beq}) 
reduce to the relation
\qq
J_{_L}\ =\ -\m J_{_R}\qquad\ {\rm on}\qquad\da M
\label{jbc}
\qqq
where, as before, $\,J_{_L}={{i\m k}}\,g\m\da_+g^{-1}$ 
and $\,J_{_R}={{i\m k}}\,g^{-1}\da_{_-}g\m$. 
Eqs.\,\,(\ref{jbc}) are the starting point of the usual
approach to the boundary WZW theory \cite{KO,Schom}. 
We preferred, however, to start from the conditions (\ref{bc}) 
and the action functional because this will allow to determine
the canonical structure of the boundary WZW model by
following a well defined procedure. This procedure 
generalizes the approach sketched in the previous section
to the case of space-times with boundary, see below.
\vskip 0.3cm

In terms of the decomposition (\ref{dem}) of the classical solutions 
into the chiral components, still implied by the bulk equation 
(\ref{ceq}), the boundary equation (\ref{jbc}) is easily
seen to be equivalent to the conditions
\qq
g_{_L}(y+2\pi)\ =\ g_{_L}(y)\,\gamma\quad\qquad{\rm and}
\quad\qquad
g_{_R}(y)\ =\ g_{_L}(-y)\,h_{_0}^{-1}\,\m\qquad\ \m\ \ \ \,\,
\label{bc0}
\qqq
which require that the chiral components be twisted-periodic
and linked to each other. Note that, by themselves, these 
relations assure that the solution given by Eq.\,\,(\ref{dem})
takes values in fixed conjugacy classes on the boundary
since they imply that
\qq
\qquad g(t,0)\ =\ g_{_L}(t)\,h_{_0}
\, g_{_L}(t)^{-1}\qquad{\rm and}\qquad
\,g(t,\pi)\ =\ g_{_L}(t-\pi)\,\gamma\, h_{_0}\, g_{_L}
(t-\pi)^{-1}.\,\, 
\label{bc1}
\qqq
The boundary conditions (\ref{bc}) determine 
these conjugacy classes. We infer that 
\qq
h_{_0}\ \in\ \CC_{_{\mu_{_0}}}\quad\ {\rm and}\ \quad 
h_{_\pi}\equiv\gamma\,h_{_0}\ \in\ \CC_{_{\mu_{_\pi}}}\,.
\label{monr}
\qqq
Consequently, the classical solutions on the strip take 
the form
\qq
g(t,x)\ =\ g_{_L}(t+x)\m\,h_{_0}\, g_{_L}(t-x)^{-1}\ =\ 
g_{_L}(t+x-2\pi)\m\,h_{_\pi}\, g_{_L}(t-x)^{-1}\,,
\label{glgl}
\qqq
where $h_{_0},\,h_{_\pi}$ and the monodromy $\gamma$ of $\m g_{_L}$ 
are constrained by the relations (\ref{monr}). 
\vskip 0.3cm

Let us denote by $\,\CP_{_{\mu_{_0}\mu_{_\pi}}}\,$ the space
of such classical solutions. It forms the phase space of the
WZW theory on the strip. As for the case of the cylinder,
$\m\CP_{_{\mu_{_0}\mu_{_\pi}}}\m$ possesses the canonical 
symplectic structure given by the symplectic form
\qq
\Omega_{_{\mu_{_0}\mu_{_\pi}}}(\delta_1\Phi,
\,\delta_2\Phi)\ =\ 
\int\limits_{M_t}\Phi^*(\iota_{_{\delta_2\Phi}}
\,\iota_{_{\delta_1\Phi}}\,d\alpha)
\ -\ \int\limits_{\da M_t}\Phi^*(\iota_{_{\delta_2\Phi}}
\,\iota_{_{\delta_1\Phi}}\,\alpha)\,.
\label{Omb}
\qqq
The boundary term is necessary to render the right hand side           
$t$-independent. Explicitly,
\qq
\Omega_{_{\mu_{_0}\mu_{_\pi}}}\ =\ {_k\over^{4\pi}}
\Big\{\int_{_0}^{^{\pi}}\hspace{-0.2cm}\tr\,\,[\m-\delta(g^{-1}
\da_tg)\,g^{-1}\delta g\,+\,2\,(g^{-1}\da_{_+}g)\,(g^{-1}
\delta g)^2\m]\,\,dx\m\,\,\cr 
+\ \omega_{_{\mu_{_0}}}(g(t,0))\ -\ \omega_{_{\mu_{_\pi}}}
(g(t,\pi))\m\Big\}
\label{Omegb}
\qqq
and its closedness is guaranteed by Eq.\,\,(\ref{beta}).
\vskip 0.3cm

The loop group elements $h\in LG$ act naturally on the space of
fields $g$ which satisfy the boundary conditions (\ref{bc}) by
\qq
g(t,x)\ \,\longmapsto\,\ h(x^+)\,\,g(t,x)\,\,
h(-x^-)^{-1}\,.
\label{blg}
\qqq
It is easy to see (for example, from the general form (\ref{glgl})
of the solutions) that they map classical solutions to classical 
solutions. The resulting action of $LG$ on $\m\CP_{_{\mu_{_0}
\mu_{_\pi}}}$ preserves the symplectic structure. In fact the choice 
(\ref{omegl}) is imposed by requiring these properties of the action 
(\ref{blg}). On the infinitesimal level the $LG$-action is generated 
by the current
\qq
J(t,x)\ =\ \cases{\hbox to 3cm{$\quad J_{_L}(t,x)$\hfill}{\rm for}
\ \ 0\leq x\leq\pi\,,\cr
\hbox to 3cm{$-J_{_R}(t,2\pi-x)$\hfill}{\rm for}\ \ \pi\leq x\leq 
2\pi\,\,}
\label{curb}
\qqq
which may be viewed as a periodic function of $x^+$ with period 
$2\pi$. Similarly, the diffeomorphisms $D\in Diff_{_+}S^1$ act on 
the space $\m\CP_{_{\mu_{_0},\mu_{_\pi}}}$ by
\qq
g(t,x)\ \,\longmapsto\,\ g_{_L}(D^{-1}(t+x))\,\,
h_{_0}\m\,g_{_L}(D^{-1}(t-x))^{-1}
\qqq
if $g$ is given by Eq.\,\,(\ref{glgl}). The action preserves
the symplectic form. It is generated infinitesimally 
by the energy-momentum tensor \,$\,T(t,x)={_1\over^{2k}}\,\tr\,J(t,x)^2$, 
again a periodic function of $x^+$ with period $2\pi$. As we see,
the WZW theory on the strip defined as above conserves half of 
the infinite-dimensional symmetries of the theory on the cylinder.
\vskip 0.3cm

In terms of the field $g_{_L}$ that parametrizes the classical
solutions via Eq.\,\,(\ref{glgl}), the symplectic form
(\ref{Omegb}) becomes
\qq
\Omega_{_{\mu_{_0}\mu_{_\pi}}}\ =\ 
\,{_k\over^{4\pi}}\Big[\int_{_0}^{^{2\pi}}
\hspace{-0.1cm}\tr\,\,(g_{_L}^{-1}\delta g_{_L})\,\,\da_x(g_{_L}^{-1}
\delta g_{_L})\m\, dx\ +\ \tr\,\,(g_{_L}^{-1}\delta g_{_L})(0)
\,(\delta\gamma)\,\gamma^{-1}\ \cr
+\ \tr\,\,(\delta h_{_0})\m h_{_0}^{-1}\,
(\gamma^{-1}\delta\gamma)\ +\ \omega_{_{\mu_{_0}}}
(h_{_0})\ -\ \omega_{_{\mu_{_\pi}}}(\gamma\,h_{_0})\Big]\,.
\label{Omll}
\qqq
Note a vague resemblance to the modified chiral symplectic
form $\tilde\Omega_{_L}$ discussed in the preceding section.
It is even more instructive to rewrite the form 
$\m\Omega_{_{\mu_{_0}\mu_{_\pi}}}$ in terms of the
vertex-IRF parametrization (\ref{dec0}) of the twisted periodic
field $g_{_L}$ which results in the decomposition
\qq
g(t,x)\ =\ =\ h_{_L}(t+x-2\pi)\,\,U\,\,
h_{_L}(t-x)^{-1}
\label{do0pi}
\qqq
of the classical solutions on the strip with
the boundary conditions (\ref{bc}), where
\qq
U\ =\ \ee^{2\pi i\m\tau}\,g_{_0}^{-1}\m h_{_0}\, g_{_0}\ =\ 
g_{_0}^{-1}\m h_{_\pi}\, g_{_0}\m,
\label{u0pi}
\qqq
combines the non-diagonal monodromy, see Eq.\,\,(\ref{glgl}). 
Inserting the parametrization
(\ref{dec0}) to (\ref{Omll}), we obtain:
\qq
&&\Omega_{_{\mu_{_0}\mu_{_\pi}}}\ =\  
{_k\over^{4\pi}}\int_{_0}^{^{2\pi}}\hspace{-0.1cm}\tr\,\,
[\m(h^{-1}{\delta}h)\,\da_x(h^{-1}{\delta}h)\,+\, 2i\m\tau\,
(h^{-1}{\delta}h)^2\,-\,2i\,({\delta}\tau)
(h^{-1}{\delta}h)\m]\m\, dx\cr\cr
&&+\,{_k\over^{4\pi}}\Big\{\tr\,(h_{_0}^{-1}\delta h_{_0})
(h_{_\pi}^{-1}\delta h_{_\pi})\,+\,\omega_{_{\mu_{_0}}}(h_{_0})\,-\,
\omega_{_{\mu_{_\pi}}}(h_{_\pi})\,+\,\omega_{_{\tau}}(\gamma)
\,+\,4\pi i\,\m\tr(\delta\tau)\m(g_{_0}^{-1}\delta g_{_0})\Big\}\cr\cr
&&\equiv\ \Omega^{LG}\ +\ \Omega^{bd}\,.
\label{tde}
\qqq
where $\,\gamma=g_{_0}\m\ee^{\m2\pi i\m\tau}\m g_{_0}^{-1}
=h_{_\pi} h_{_0}^{-1}$.
Above $\omega_{_\tau}(\gamma)$ is really a form depending
on the pair $(g_{_0},\tau)$ rather than on $\gamma$ since 
the latter, unlike $h_{_0}$ and $h_{_\pi}$, is not restricted to a single 
conjugacy class. We recognize the symplectic form $\m\Omega^{LG}$ 
of the loop group model space $\CM_{_{LG}}$ as the first part of 
$\m\Omega_{_{\mu_{_0}\mu_{_\pi}}}$. \,The next section 
is devoted to the interpretation of the second
part $\m\Omega^{bd}$ involving the boundary data 
$(h_0,h_\pi,g_0,\tau)$ defined modulo simultaneous adjoint
action of $\,G$ on $\,h_{_0}$ and $\,h_{_\pi}$ and left
action on $\,g_0$.
\vskip 0.9cm

\nsection{Relation to the Chern-Simons theory}
\subsection{CS theory on a sphere}
\vskip 0.5cm

The 2-form $\m\Omega^{bd}$ may be identified with the 
symplectic structure on the phase space of the CS
theory on $S^2\times\NR$ with three Wilson lines. 
We shall represent $S^2$ as the complex projective plane 
$\,\NC P^1\cong\NC\cup\{\infty\}\,$ and we shall fix
3 punctures on it, say at points points $0$, $\pi$ and 
$w\equiv\hf\pi+2i$, see Figs.\,\,3.

\leavevmode\epsffile[-95 -25 225 300]{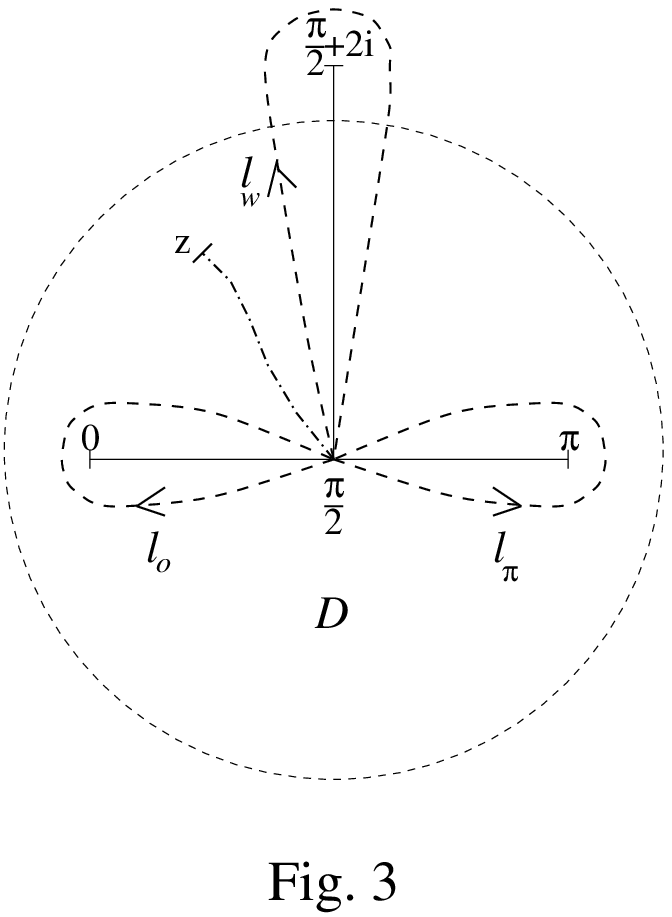}

\noindent Let us consider flat unitary gauge potentials 
(connections) $A$ on $\,\NC P^1\setminus\{0,\pi,w\}$ with values 
in the Lie algebra $\Ng$ and such that
\qq
A\ =\ \cases{\hbox to 5.6cm{$-\eta_{_0}\m\mu_{_0}\m
\eta_{_0}^{-1}\,\,\m{1\over{2i}}({{dz}\over{z}}
-{{d\bar z}\over{\bar z}})$
\hfill}{\rm around}\ \ \ 0\,,\cr\cr
\hbox to 5.6cm{$\ \,\m\eta_{_\pi}\mu_{_\pi}
\eta_{_\pi}^{-1}\,{1\over{2i}}({{dz}\over{z-\pi}}
-{{d\bar z}\over{\bar z
-\pi}})$\hfill}{\rm around}\ \ \ \pi\,,\cr\cr
\hbox to 5.6cm{$-\eta_{_w}\m\tau\,\eta_{_w}^{-1}\,\,\,
{1\over{2i}}({{dz}\over{z-\hf\pi-2i}}-{{d\bar z}
\over{\bar z-\hf\pi+2i}})$\hfill}{\rm around}\ \ \ w\,,}
\label{assy}
\qqq
where $\,\eta_{_0},\,\eta_{_\pi},\,\eta_{_w}\m\in\m G\m$ and 
where $\m\mu_{_0}$, $\mu_{_\pi}$ and $\tau$ are, as before, 
in the positive Weyl alcove $\m\CA_{_W}\subset\Nt\m$, 
the first two fixed and the last arbitrary. 
The closed 2-form on the infinite-dimensional space
of flat gauge potentials $A$ with the behavior (\ref{assy})
around the punctures,
\qq
\Omega^{CS}\ =\ -{_k\over^{4\pi}}
\int\limits_{\NC}\tr\,\,(\delta A)^2\ +\ k\m i\,\m\tr\,[
\m-\,\mu_{_0}\m(\eta_{_0}^{-1}\delta\eta_{_0})^2\ +\ 
\mu_{_\pi}\m(\eta_{_\pi}^{-1}\delta\eta_{_\pi})^2\,\cr
\ -\ \tau\m(\eta_{_w}^{-1}\delta\eta_{_w})^2
\ +\ (\delta\tau)\m(\eta_{_w}^{-1}\delta\eta_{_w}\m]\,,
\label{OmCS}
\qqq
is invariant under the gauge transformations $\,h:\NC P^1
\rightarrow G$ constant around the punctures acting on the 
gauge potentials by
\qq
A\ \ \longmapsto\ \ h\m A\, h^{-1}\ -\ i\,(dh) h^{-1}\,.
\qqq
It descends to the quotient space $\,\CP^{CS}_{_{\mu_{_0}\mu_{_\pi}}}$ 
making it a finite-dimensional symplectic manifold that may be 
identified as the phase space of the CS theory on 
$\NC P^1\times\NR$ with timelike Wilson lines passing through 
the punctures, see Fig.\,\,4.

\leavevmode\epsffile[-110 -20 215 305]{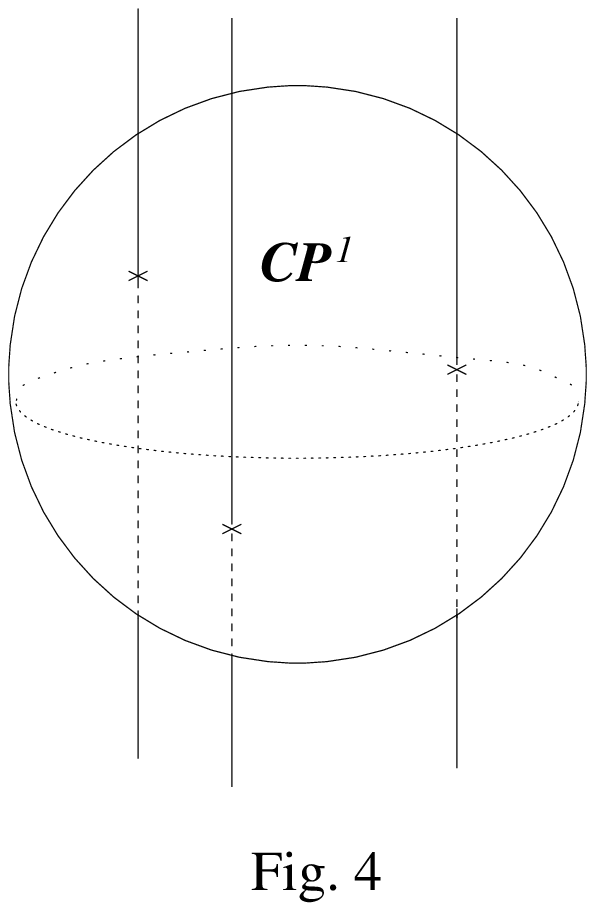}

\noindent We shall keep the notation $\m\Omega^{CS}$ 
for the symplectic form on $\m\CP^{CS}_{_{\mu_{_0}\mu_{_\pi}}}$. 
As before, the whole construction may be repeated in the complex 
setup where we relax unitarity of the connections and end up with 
a complex symplectic manifold.
\vskip 0.3cm

The symplectic forms on the phase spaces of the CS 
theory on general punctured Riemann surfaces have been explicitly 
computed in terms of the holonomy of flat connections
in the reference \cite{AlMal}, see Theorem 1 therein. 
The idea of that computation is simple. One renders the surface 
simply connected by cutting it (in our case along the
inverted letter $T$ in Fig.\,\,3). 

On the cut surface, any flat connection $A$ is pure gauge so that 
if one defines
\qq
g(z)\ =\ {\mathop{\ee}\limits^{_\leftarrow}}^{\,\,i
\int\limits_{\ell_z}
\hspace{-0.1cm}A}
\label{gocs}
\qqq
for any path $\ell_z$ in the cut surface connecting the base point 
($\hf\pi-0+i0$ in our case) to $z$, where 
$\m{\mathop{\ee}\limits^{_\leftarrow}}
\m$ denotes the path-ordered exponential, then $\,A={_1\over^i}
\m(dg)g^{-1}$. \,The identity
\qq
\tr\,\,(\delta A)^2\ =\ -\m d\,[\m\tr\,(g^{-1}\delta g)\m 
d(g^{-1}\delta g)\m]
\label{tibp}
\qqq
permits to replace the surface integral in the definition 
(\ref{OmCS}) by the integral along the boundary of the cut 
surface which forms a contour that may be decomposed 
into the generators of the fundamental group. The rest 
of the argument is a straightforward, although tedious, 
bookkeeping. 
\vskip 0.3cm

For the case at hand with three punctures in the complex 
projective plane, the local behavior (\ref{assy}) 
around the punctures assures that the holonomy of $A$ takes 
values in prescribed conjugacy classes:
\qq
h_{_0}\,\equiv\,
{\mathop{\ee}\limits^{_\leftarrow}}^{\,\,i\int\limits_{\ell_0}
\hspace{-0.1cm}A}\ \in\ \CC_{_{\mu_{_0}}}\,,\quad
h_{_\pi}\,\equiv\,
{\mathop{\ee}\limits^{_\leftarrow}}^{\,i\int\limits_{\ell_\pi}
\hspace{-0.1cm}A}\ \in\ \CC_{_{\mu_{_\pi}}}\,,\quad
\gamma\,\equiv\,
{\mathop{\ee}\limits^{_\leftarrow}}^{\,i\int\limits_{\ell_w}
\hspace{-0.1cm}A}\ \in\ \CC_{_{\mu_{_\tau}}}\,,
\label{holo}
\qqq
where $h_{_0}$, $h_{_\pi}$ and $\gamma=h_{_\pi} h_{_0}^{-1}$ denote now 
the parallel transporters in the gauge potential $A$
along the closed paths $\ell_0$, $\ell_\pi$ and 
$\ell_w$ starting at $\hf\pi$, see Fig.\,\,3. Writing 
$\,\gamma=g_{_0}\m\ee^{\m2\pi i\m\tau}g_{_0}^{-1}$, one obtains 
the identification 
\qq
\CP^{CS}_{_{\mu_{_0}\mu_{_\pi}}}\ \ \cong\ \ 
\Big\{\,(h_{_0},\m h_{_\pi},\m g_{_0},\m\tau)\in\m\CC_{_{\mu_{_0}}}
\times\CC_{_{\mu_{_\pi}}}\times G\times\CA_{_W}\,\Big\vert 
\m\, h_{_\pi} h_{_0}^{-1}=g_{_0}\m\ee^{2\pi i\m\tau}g_{_0}^{-1}\,\Big\}
\Big/\m G\,.
\label{pcs}
\qqq
The simultaneous adjoint action of $G$ on $h_{_0}$ and $h_{_\pi}$
and the left action on $g_{_0}$, whose orbit space is taken above,
is induced on the holonomy by the local gauge transformations 
of the gauge potentials $A$. It appears then that, expressed in 
the language of $\,(h_{_0},\m h_{_\pi},\m g_{_0},\m\tau)\m$, \,the symplectic
form $\m\Omega^{CS}$ on $\,\CP^{CS}_{_{\mu_{_0}\mu_{_\pi}}}\m$,
see (\ref{OmCS}), coincides with the $\m\Omega^{bd}$ part 
of the symplectic form $\m\Omega_{_{\mu_{_0}\mu_{_\pi}}}$
given by Eq.\,\,(\ref{tde}) on the phase space of the boundary 
WZW model. Note the symplectic action of the Cartan subgroup 
$T$ on $\,\CP^{CS}_{_{\mu_{_0}\mu_{_\pi}}}$ by 
$\,g_{_0}\mapsto g_{_0}t^{-1}$.
\vskip 0.3cm

The phase space $\,\CP_{_{\mu_{_0}\mu_{_\pi}}}$ of the boundary 
WZW theory may be viewed as the symplectic reduction 
with respect to the diagonal action of the Cartan subgroup $T$
of the product of the model space $\CM_{LG}$ for the loop group 
and of $\,\CP^{CS}_{_{\mu_{_0}\mu_{_\pi}}}$:
\qq
\CP_{_{\mu_{_0}\mu_{_\pi}}}\ \cong\ 
(\CM_{_{LG}}\,\times\CP^{CS}_{_{\mu_{_0}\mu_{_\pi}}})\,\Big/
\hspace{-0.15cm}\Big/\,T
\ =\ (\CM_{_{LG}}\,\times_{_{\CA_{_W}}}
\CP^{CS}_{_{\mu_{_0}\mu_{_\pi}}})\,\Big/\,T\,\m,
\label{strbd}
\qqq
where the fiber product over $\CA_{_W}$ equates the $\tau$ components 
of $\,\CM_{_{LG}}$ and of $\m\CP_{_{\mu_{_0}\mu_{_\pi}}}$.
This is the main structural result of thhis subsection, which
should be compared with the preceding result (\ref{strL}) 
about the structure of the chiral phase space of the WZW theory 
on the cylinder.
\vskip 0.3cm

As is well known, the phase space of the CS
theory on a punctured Riemann surface, with the holonomies
around the punctures constrained to the conjugacy classes 
$\CC_{_{\lambda_i/k}}$ where $\lambda_i$ are weights, may be 
quantized. Upon quantization it gives rise to the 
finite-dimensional space $\, W_{_{k,\m(\lambda_i)}}\,$ 
of the conformal blocks of the WZW theory with insertions 
of primary fields in representations of $G$ with highest 
weights $\lambda_i$. Consequently, the classical decomposition 
(\ref{strbd}) suggests the following realization 
of the quantum space of states of the WZW theory 
with the boundary conditions (\ref{bc}) if $\m\mu_{_0}
=\lambda_{_0}/k$ and $\mu_{_\pi}=\lambda_{_\pi}/k$ where $\lambda_{_0}$ 
and $\lambda_{_\pi}$ are weights:
\qq
\CH_{_{\lambda_{_0}\lambda_{_\pi}}}\ =\ \mathop{\oplus}\limits_\lambda
\,\, \CV_{_{k,\lambda}}\otimes\m W_{_{k,\m\lambda_{_0}\bar{\lambda}_\pi
\m\lambda}}\,\m,
\label{debd}
\qqq
where the sum is over the weights $\lambda$ with $\lambda/k$
in the positive Weyl alcove $\CA_{_W}$. 
By definition, $\bar{\mu}$ labels the conjugacy 
class inverse to $\m\CC_{_{\mu}}$ and $\m\bar\mu_{_\pi}=\bar
\lambda_{_\pi}/k$. The replacement of $\mu_{_\pi}$ by $\bar\mu_{_\pi}$
is due to the opposite orientation of the contour $\ell_\pi$ 
in Fig\,\,3. 
\vskip 0.3cm

The decomposition (\ref{debd}) is consistent with results 
of the general theory of conformal boundary conditions
\cite{Card,CardL1}. That theory states that, for 
the so called diagonal models whose examples are provided
by the WZW theories with simply connected groups, the boundary 
conditions are in a one-to-one correspondence with the primary 
fields of the bulk model. Indeed, in our case\footnote{See 
\cite{Istam} for more details on how quantization chooses 
the discrete family of conjugacy classes.}, \,both are labeled 
by the weights $\lambda$ in $k\CA_{_W}\subset \Nt$.
Moreover, the general theory asserts that the irreducible 
representations of the chiral algebra (in our case, 
of the Kac-Moody algebra) appear in the boundary theory Hilbert 
spaces with the multiplicities given by the (Verlinde) dimensions 
of the spaces of 3-point conformal blocks 
\qq
N_{_{\lambda_{_0}\lambda}}^{^{\,\lambda_{_\pi}}}\ =\ {\rm dim}\,\,
W_{_{k,\m\lambda_{_0}\bar{\lambda}_{_\pi}\lambda}}\,,
\qqq
in agreement with the decomposition (\ref{debd}).
As we shall see below, our classical results allow, 
however, for more. They permit, for example, to quantize 
naturally the basic fields of the boundary WZW model. 
\vskip 0.6cm

\subsection{CS theory on a disc}
\vskip 0.2cm

The result of the last subsection permits to establish an even
more direct relation between the boundary WZW model
and the CS theory on a 3-manifold with boundary.
Let us consider the CS theory on $D\times \NR$ where
$D$ is a disc of radius 1 centered at $\hf\pi$, see Fig.\,\,3, 
with two timelike Wilson lines passing through the punctures 
at $0$ and $\pi$. The phase space $\,\CP^{CS}_{_{\hspace{-0.1cm}D,
\mu_0\mu_\pi}}\,$ of the theory is composed of flat connections 
$\,A_{_D}$ on $\,D\,$ with the representation as in (\ref{assy}) 
around $0$ and $\pi$, modulo gauge transformation constant 
around the punctures and equal to $1$ on $\,\partial D$. 
\,The symplectic form $\,\Omega^{CS}_{_D}\,$ is given by the
first line of (\ref{OmCS}) with the integral restricted to
$\,D$. The phase space $\,\CP^{CS}_{_{\hspace{-0.1cm}
D,\mu_0\mu_\pi}}\,$
may be easily identified with the phase space 
$\,\CP_{_{\mu_0\mu_\pi}}\,$ of the boundary WZW model using 
the map
\qq
A_D\ \longmapsto\ (h_0,h_\pi,\tau,\,h)\,,
\label{mapC}
\qqq
where $\,h_{_0}$ and $\,h_{_\pi}$ are defined as in (\ref{holo})
and describe the holonomy of $\,A_{_D}$ around the punctures,
$\,h_{_\pi}h_{_0}^{-1}=g_{_0}\m{\rm e}^{\,2\pi i\,\tau}\,g_{_0}^{-1}$,
\,and
\qq
h(x)=g(\hf\pi+i\,{\rm e}^{\,i\,x})\,g_0\,{\rm e}^{-i\,\tau\,x}\,,
\label{loop}
\qqq
with $\,g(z)\,$ given by (\ref{gocs}). It is easy to see that
$\,h\,$ is periodic, i.e.\,\,that it belongs to the loop group.
The gauge transformations of $\,A_{_D}$ induce a simultaneous
adjoint action of $\,G$ on the holonomy $\,h_{_0}$ and $\,h_{_\pi}$
and do not change $\,h$. The map of the orbits is 1 to 1
and establishes the isomorphism between 
$\,\CP^{CS}_{_{\hspace{-0.1cm}D,\mu_0\mu_\pi}}$
and $\,\CP_{_{\mu_0\mu_\pi}}$. \,It remains to identify the symplectic 
structures of two phase spaces. This may be done along the lines
of \cite{AlMal} or using directly the result of that reference. 
In the latter case, we extend a connection $\,A_{_D}$ on $\,D\,$ 
to a flat connection $\,A\,$ on $\,\NC P^1$ with three punctures,
with the behavior (\ref{assy}) around them, and write
\qq
\Omega^{CS}_{_D}\ =\ \Omega^{CS}\ +\ {_k\over^{4\pi}}
\int\limits_{\NC\setminus D}\tr\,\,(\delta A)^2\ +\ k\m i\,\m\tr\,[
\m\tau\m(\eta_{_w}^{-1}\delta\eta_{_w})^2
\ -\ (\delta\tau)\m(\eta_{_w}^{-1}\delta\eta_{_w}\m]\,,
\qqq
see (\ref{OmCS}). As we have discussed, $\,\Omega^{CS}\,$
reproduces the boundary part $\,\Omega^{\rm bd}\,$ of
the symplectic form $\,\Omega_{_{\mu_0\mu_\pi}}$ of
$\,\CP_{_{\mu_0\mu_\pi}}$, \,see (\ref{tde}). It remains 
to see that the other term reproduces $\Omega^{LG}$.
This is an old result \cite{EMSS} which says that the CS
phase spaces on a disc with one puncture may be identified
with the coadjoint orbits of $\,\hat{LG}$. It may be
easily established using (\ref{tibp}) and integrating
by parts. 
\vskip 0.3cm

The relation of the the boundary WZW theory to the CS theory 
on twice punctured disc is certainly worth pursuing further. 
As mentioned in Introduction, it may lead to new applications of
the boundary theory. It is also a source of natural structures 
in the boundary models that are less visible in the original
formulation. It also raises a natural question about the 
interpretation of the CS theory on a disc with more than two 
punctures. 
\vskip 0.9cm

\nsection{Relation to the Poisson-Lie groups}
\vskip 0.5cm

Reference \cite{AlMal} contains another valuable result,
stated in Theorem 2 therein.
It realizes the (complex versions) of the CS
theory phase spaces in Poisson-Lie terms. Let us recall 
how this is done. Consider the product space $\,\CC_{_{\mu_{_0}}}
\times\CC_{_{\bar\mu_{_\pi}}}\times\CM_{_G}^{PL}\,$  
composed of the elements $\,(\gamma_{_0},\gamma_{_\pi},(\sigma_{_w},\tau))\,$
with  
\qq
\gamma_{_0}&=&\hbox to 2.4cm{$\sigma_{_0}\m\ee^{\m2\pi i\m\mu_{_0}}\,
\sigma_{_0}^{-1}$\hfill}\ =\ \gamma_{_{0-}}\gamma_{_{0+}}^{-1}\,,\cr 
\gamma_{_\pi}&=&\hbox to 2.4cm{$\sigma_{_\pi}\m\ee^{-2\pi i\m\mu_{_\pi}}
\,\sigma_{_\pi}^{-1}$\hfill}\ =\ \gamma_{_{\pi-}}\gamma_{_{\pi+}}^{-1}\,,\\
\gamma_{_w}&=&\hbox to 2.4cm{$\sigma_{_w}\m\ee^{\m2\pi i\m\tau}\,
\sigma_{_w}^{-1}$\hfill}\ =\ \gamma_{_{w-}}\gamma_{_{w+}}^{-1}\,.
\nonumber
\qqq
Recall that the Poisson-Lie model space $\m\CM_{_G}^{PL}$ 
comes equipped with the symplectic form $\m\Omega^{PL}$,
see Eq.\,\,(\ref{deco}). In turn, upon fixing $\tau$,
$\m\Omega^{PL}$ induces the symplectic forms $\m\Omega^{PL}_\tau$ 
on the conjugacy classes $\CC_{_\tau}$ which are identified 
with the symplectic leaves of the dual Poisson-Lie group $G^*$. 
The product space $\,\CC_{_{\mu_{_0}}}\times\CC_{_{\bar\mu_{_\pi}}}
\times\CM_{_G}^{PL}\,$ may be equipped with the symplectic
structure 
\qq
\Omega^{PL}_{\mu_{_0}}\,+\,\m\Omega^{PL}_{\bar
\mu_{_\pi}}\,+\,\Omega^{PL}\,.
\label{comp}
\qqq
Define now the map
\qq
\CC_{_{\mu_{_0}}}\times\CC_{_{\bar\mu_{_\pi}}}\times
\CM_{_G}^{PL}\,\ni\,(\gamma_{_0},\gamma_{_\pi},(\sigma_{_w},\tau)) 
\ \mapsto\ (h_{_0},h_{_\pi},(g_{_0},\tau))
\,\in\,\,\CC_{_{\mu_{_0}}}\times\CC_{_{\bar\mu_{_\pi}}}\times 
G\times\CA_{_W}
\qqq
by setting 
\qq
h_{_0}\ =\ \gamma_{_0}\,,\quad\ h_{_\pi}^{-1}\ =\ \gamma_{_{0+}}
\gamma_{_\pi}\gamma_{_{0+}}^{\m-1}\,,\quad\ g_{_0}\ =\ \gamma_{_{0+}}
\gamma_{_{\pi+}}\sigma_{_w}\,.
\label{mapt}
\qqq
Note that if we set $\,\gamma=g_{_0}\m\ee^{\m2\pi i\m\tau}g_{_0}^{-1}\m$
then $\,\gamma=\gamma_{_{0+}}\gamma_{_{\pi+}}
\gamma_{_w} \gamma_{_{\pi+}}^{-1}\gamma_{_{0+}}^{-1}\m.$
\,The separate Poisson-Lie action of $G$
on $\m\CC_{_{\mu_{_0}}}$, $\m\CC_{_{\bar\mu_{_\pi}}}$
and $\m\CM_{_G}^{PL}$ given by the adjoint action on $\gamma_0$
and $\gamma_\pi$ and the left action on $\sigma_{_w}$ has a 
(twisted-)diagonal version. This version consists of the simultaneous 
adjoint action of $G$ on $h_{_0}$, and $h_{_\pi}$ and the left action 
on $g_{_0}$ and may be used to perform a Poisson-Lie version 
of the symplectic reduction of the product manifold 
$\,\CC_{_{\mu_{_0}}}\times\CC_{_{\bar\mu_{_\pi}}}\times 
\CM_{_G}^{PL}$. In concrete terms, the reduction imposes 
the condition
\qq
\gamma_{_{0-}}\gamma_{_{\pi-}}\gamma_{_{w-}}\ =\ 1\ =\ 
\gamma_{_{0+}}\gamma_{_{\pi+}}\gamma_{_{w+}} 
\qqq
which is the same as $\m h_{_\pi} h_{_0}^{-1}
=\gamma\m$ and passes to the space of orbits of the (twisted-)diagonal 
Poisson-Lie action of $\m G$. The symplectic form (\ref{comp}) descends 
to the reduced space. Recalling from the previous section 
the realization (\ref{pcs}) of the phase space of the CS 
theory on the projective plane with three punctures, we infer that 
the map (\ref{mapt}) induces the isomorphism\footnote{More exactly,
an isomorphism between open dense subspaces.}
\qq
\CP^{CS}_{_{\mu_{_0}\mu_{_\pi}}}\ \ \ \cong\ \ \ 
(\CC_{_{\mu_{_0}}}\times\CC_{_{\bar\mu_{_\pi}}}\times
\CM_{_G}^{PL})\,\Big/\hspace{-0.15cm}\Big/\,G\,.
\label{prese}
\qqq
A direct calculation \cite{AlMal} shows then that this is 
an isomorphism of symplectic manifolds.
\vskip 0.3cm

Note that in terms of the parametrization (\ref{mapt}), the monodromy
part (\ref{u0pi}) in the decomposition (\ref{do0pi}) of the classical
solutions of the boundary theory take the form
\qq
U\ =\ \sigma_{_w}^{-1}\m\gamma_{_{\pi-}}^{-1}\m
\gamma_{_{\pi+}}\m\sigma_{_w}\,.
\label{us}
\qqq
Below, we shall quantize these expressions. This will permit 
an explicit construction of the action of the quantum bulk fields 
$\,g(t,x)\m$ in the spaces of states of the boundary WZW theory.
\vskip 0.9cm

\nsection{Quantization of the boundary theory}
\subsection{The space of states}
\vskip 0.3cm

The isomorphism (\ref{prese}) has its counterpart at the
quantum level which allows for another presentation of the
space of states of the boundary theory, see (\ref{debd}). 
Under quantization, the symplectic space
$\m\,\CC_{_{\mu_{_0}}}\times\CC_{_{\mu_{_\pi}}}\times
\CM_{_G}^{PL}\,$ for $\m\mu_{_{0,\pi}}=\lambda_{_{0,\pi}}/k\m$ 
becomes $\,{\mathop{\oplus}\limits_\lambda}\,\CV_{_{q,\m\lambda_{_0}
\bar\lambda_{_\pi}\lambda}}\,$ in the shorthand notation 
$\,\CV_{_{q,\m(\lambda_i)}}
\equiv\otimes_{_i}\CV_{_{q,\m\lambda_i}}\m$ for the tensor
product of the highest weight representations of the deformed
enveloping algebra $\,\CU_q(\Ng)$.  \,The diagonal 
Poisson-Lie action of $G$ gives rise on the quantum level 
to the diagonal (coproduct induced) action of $\,\CU_q(\Ng)$ 
in the latter space. In the first approximation, 
the subspace $\,\mathop{\oplus}\limits_\lambda\,\,
\CV^{^{\,inv}}_{_{q,\m\lambda_{_0}\bar\lambda_{_\pi}\,\lambda}}\,$ 
of the invariant tensors of that action gives the space of states 
corresponding to the symplectic reduction $\,(\CC_{_{\mu_{_0}}}\times
\CC_{_{\bar\mu_{_\pi}}}\times\CM_{_G}^{PL})\,\Big/\hspace{-0.15cm}
\Big/\,G\m.$ \,More precisely, the subspaces of invariants 
$\,\CV^{^{\,inv}}_{_{q,\m(\lambda_i)}}\subset\CV_{_{q,\m(\lambda_i)}}\,$ 
may be equipped with a semi-positive scalar product (coming from natural 
hermitian forms on the spaces $\m\CV_{_{q,\lambda}}$) and one should 
divide by the subspaces of null-vectors. The quotient spaces 
$\,\CW_{_{q,\m(\lambda_i)}}\,$ are isomorphic 
to the spaces $\CW_{_{k,\m\lambda_{_0}\bar\lambda_{_\pi}\,\lambda}}$ 
of the conformal blocks of the WZW theory. Consequently, we obtain
the following presentation of the space of states of the boundary WZW 
theory defined in Eq.\,\,(\ref{debd}):
\qq
\CH_{_{\lambda_{_0}\lambda_{_\pi}}}\ \cong\ \ \mathop{\oplus}\limits_\lambda
\,\,\CV_{_{k,\lambda}}\otimes\CW_{_{q,\m\lambda_{_0}
\bar\lambda_{_\pi}\lambda}}\ \subset\ \CH_{_L}\otimes
\CH^{\,bd}_{_{\lambda_{_0}\lambda_{_\pi}}}\,,
\label{ndebd}
\qqq
where the chiral space of states $\,\CH_{_L}$ is given by
Eq.\,\,(\ref{chss}) and
\qq
\CH^{\,bd}_{_{\lambda_{_0}\lambda_{_\pi}}}\ =\ \mathop{\oplus}
\limits_\lambda\,\,\CW_{_{q,\m\lambda_{_0}
\bar\lambda_{_\pi}\lambda}}\,.
\label{inv0}
\qqq
This realization of the space $\,\CH_{_{\lambda_{_0}\lambda_{_\pi}}}$ 
will allow to define the action of the quantized bulk fields 
$\,g(t,x)\m$ in $\,\CH_{_{\lambda_{_0}\lambda_{_\pi}}}$ 
by finding the quantum 
version of the decomposition (\ref{do0pi}) of the classical field. 
As in the bulk case, quantization of the factors $h_{_L}$ and 
$h_{_L}^{-1}$ will be given by the vertex operators of the Kac-Moody 
algebra whereas the monodromy factor $\m U$ will be realized 
by operators acting in the space $\CH^{\,bd}_{_{\lambda_{_0}
\lambda_{_\pi}}}$. In what follows, we shall carry out this construction 
in detail for the case of the group $SU(2)$.
\vskip 0.6cm

\subsection{The Kac-Moody vertex operators}
\vskip 0.2cm

For $\,G=SU(2)$, \,the weights $\,\lambda\,$ such that 
$\,\lambda/k\in\CA_{_W}\,$ are labeled by spins $j=0,\hf,\dots,
{k\over2}\,$: $\,\lambda=(\matrix{_j&_0
\cr^0&^{-j}})\m$. \,It was shown in \cite{BerFel} that one may realize 
the spaces $\CV_{_{k,j}}$ carrying the irreducible highest 
weight representations
of the Kac-Moody algebra $\hat{su}(2)$ as the cohomology of a complex
of Fock spaces $\CF_{\alpha}$. The latter carry representations of
the CCR algebra\footnote{In \cite{BerFel}, $\,\beta_n$ and $\gamma_n$
are denoted, respectively, $\omega_n$ and $\omega^\dagger_n$.}
\qq
[a_n,a_m]=n\,\delta_{n,-m}\,\ \quad [\beta_n,\gamma_m]=\delta_{n,-m}
\qqq
with all the other commutators vanishing. $\CF_\alpha$ are built
by applying the creation operators $a_n$, $\beta_{n+1}$ and $\gamma_n$
with $n<0$ to the vacuum vector $|\alpha\rangle$ s.t. $a_0\,|\alpha\rangle
\,=\,\alpha\m|\alpha\rangle\m$ and 
$\m a_n\m|\alpha\rangle=\beta_n\m|\alpha\rangle =\gamma_{n-1}\m|\alpha
\rangle=0\,$ for $n>0$. One introduces the free field vertex operators
depending on the real variable $x$, the Wick ordered exponentials
of the (chiral) free field $\,\phi(x)=\phi_{_0}+a_{_0}x+i
\sum\limits_{n\not=0}{_1\over^n}\,a_n\,\ee^{-inx}$,
\qq
\psi_\eta(x)\ =\ :\ee^{-i{_\eta\over^{2\xi}}\m\phi(x)}:
\ \equiv\ \ee^{-i{_\eta\over^{2\xi}}\m \phi_{_0}}\,
\ee^{-i{_\eta\over^{2\xi}}\m a_{_0}\m x}
\,\ee^{\m{_\eta\over^{2\xi}}\sum\limits_{n<0}{1\over n}a_n\ee^{-inx}}
\,\ee^{\m{_\eta\over^{2\xi}}\sum\limits_{n>0}{1\over n}a_n\ee^{-inx}}.
\label{vop}
\qqq
Above, $\,\xi=\sqrt{_{k+2}\over^2}\,$ and $\,\ee^{-i\m{_\eta\over^{2\xi}}
\m\phi_{_0}}|\alpha\rangle=|\alpha-{_\eta\over^{2\xi}}\rangle$
and it commutes with all the generators of the $CCR$ algebra
but $a_{_0}$. The operators $\psi_\eta(x)$ are twisted-periodic:
$\,\psi_\eta(x+2\pi)=\psi_\eta(x)\,\ee^{-{_{\pi i}\over^\xi}\m\eta
\m a_0}$ and
they satisfies the commutation relations
\qq
\psi_\eta(x)\ \psi_{\eta'}(x')\ =\ \ee^{\m {_{\pi i\m\eta\eta'}\over^{k+2}}\,
[E({x-x'\over2\pi})+{1\over2}]}\ \,\psi_{\eta'}(x')\ \psi_\eta(x)\,,
\label{comr}
\qqq
where $E(\cdot)$ denotes the ``entier'' function. Introduce also
the $\beta,\,\gamma$ fields $\,\beta(x)=\sum\beta_n\,
\ee^{-inx}$, $\,\gamma(x)=\sum\gamma_n\,\ee^{-inx}\,$
which are periodic in $x$ and which satisfy the commutation relation
\qq
[\beta(x)\m,\,\gamma(x')]\ =\ 2\pi\,\delta(x-x')\,.
\label{bgcr}
\qqq
The quantum version of the free field construction (\ref{ffr1}) of 
the current reads
\qq
J\ \equiv\ \left(\matrix{J^3&\,J^+\cr J^-&-J^3}\right)
\ =\,\left(\matrix{-\xi\,\partial\phi\,-
:\beta\m\gamma:&\ -ik\,\partial\beta\,+\,2\xi\,\beta\,
\partial\phi\,+:\beta^2\gamma:\cr-\gamma&\ \xi
\,\partial\phi\,+:\beta\m\gamma:}\right)\,.
\label{qfr1}
\qqq
It goes back to Wakimoto \cite{Wak} and may be easily rewritten
in terms of the current modes such that 
$\,J(x)=\sum J_n\,{\rm e}^{-inx}$. The quantized current satisfies 
for $\vert x-x'\vert<2\pi$ the commutation relations
\qq
[J(x)_{_1}\m,\,J(x')_{_2}]\ =\ 2\pi\,\delta(x-x')\,[J(x)_{_1}\m,
\,C_{_{12}}]\ +\ 2\pi\m i\m k\,\delta'(x-x')\,C_{_{12}}\,,
\qqq
a quantum counterpart of the Poisson bracket of the classical
current. 
The action of the quantized current turns the Fock spaces $\CF_\alpha$ 
into the modules of the $\hat{su}(2)$ affine Kac-Moody algebra. 
The (screening) operators 
\qq
Q(x)={_1\over^{ik}}\,\ee^{\m {_{\pi i}\over^\xi}\m a_{_0}}
\hspace{-0.15cm}\int\limits_x^{x+2\pi}
\gamma(y)\,\psi_{_2}(y)\,dy\,
\label{scrop}
\qqq
are nilpotent $Q(x)^{k+2}=0$ and define for $j=0,\hf,\dots,{_k\over^2}$ 
and $Q\equiv Q(x)$ an $x$-independent complex  
\qq
\dots\ \,\mathop{\longrightarrow}\limits^{Q^{2j+1}}
\,\ \CF_{k+1-j\over\xi}\ \,
\mathop{\longrightarrow}\limits^{Q^{k+1-2j}}\,\ \CF_{j\over\xi}\,\ 
\mathop{\longrightarrow}\limits^{Q^{2j+1}}\,\ \CF_{-{j+1\over\xi}}
\,\ \mathop{\longrightarrow}\limits^{Q^{k+1-2j}}\ \,\dots
\qqq
of $\hat{su}(2)$-modules whose middle cohomology 
gives the irreducible highest weight module $\CV_{_{k,j}}$ 
of $\hat{su}(2)$, see \cite{BerFel}. The energy-momentum
tensor
\qq
T\ =\ \hf:(\da\phi)^2:-\,{_1\over^{2\xi}}\,\da^2\phi\ -\,:\gamma\,\da\beta:
\label{emtt}
\qqq
satisfies the Virasoro commutation relations
\qq
[T(x)\m,\,T(x')]\ =\ -\m4\pi\m i\,\delta'(x-x')\m\,T(x')\,+\,
2\pi\m i\,\delta(x-x')\m\,\da T(x')\cr\cr
-\,{_{\pi\m i\m k}\over^{2(k+2)}}
\m\,(\delta'(x-x')+\delta'''(x-x'))
\label{Vir}
\qqq
corresponding to the value ${3k\over k+2}$ of the Virasoro
central charge.
\vskip 0.3cm

Quantization of the $SU(2)$-valued fields $h_{_L}(x)$
may be guessed from the classical free field representation
(\ref{ffr2}). In fact, the last term, the matrix involving 
$(\Pi-\Pi^{-1})^{-1}$, has to be handled with care to avoid
singularities. We find it convenient to reshuffle such terms
between the current algebra and the quantum group degrees
of freedom and to introduce four different matrices of quantum
operators, $\chi(x)$, $\tilde\chi(x)$, $u(x)$ and $\tilde u(x)$. 
The first one,
\qq
\chi\ =\ \left(\matrix{\beta\,\psi_{_{-1}}\,Q&\ 
\beta\,\psi_{_{-1}}\cr \psi_{_{-1}}\,Q&\ \psi_{_{-1}}}\right)\,.
\label{kmvo}
\qqq
should be thought of as quantization of the field $h'_{_L}$, a modified
version of $h_{_L}$ with the factor involving $(\Pi-\Pi^{-1})^{-1}$ 
on the right hand side of (\ref{ffr2}) dropped. Note that, compared 
to the classical expression (\ref{ffr2}), we have also dropped the term 
involving $\psi(x)$. This is a more delicate Wick-ordering renormalization 
effect. The components of $\chi(x)$ form the Kac-Moody vertex operators
that descend to the Fock-space cohomology, see \cite{BerFel}.
As the result, they may be viewed as operators acting in the chiral space 
of states\footnote{More exactly, they are operator-valued
distributions and map into a completion of $\CH_{_L}$.} 
\qq
\CH_{_L}\ =\ \mathop{\oplus}\limits_{^{j=\hf,\dots,{_k\over^2}}}
\hspace{-0.1cm}\CV_{_{k,j}}\,.
\label{hel}
\qqq
The components $\chi_{_{a1}}$ lower the value of $\,j$
by ${1\over2}$ and the $\chi_{_{a2}}$ ones raise it by ${1\over2}$.
\vskip 0.3cm

The other matrices of operators that we shall consider are modified 
versions of $\,\chi$. They are defined as follows:
\qq
\tilde\chi\ =\ \left(\matrix{\psi_{_{-1}}&\ -\beta\,\psi_{_{-1}}\cr
-\psi_{_{-1}}\,Q&\ \beta\,\psi_{_{-1}}\,Q}\right),\qquad\quad
u\ =\ \chi\,{_1\over^{[p]}}\,,\qquad\quad\tilde u\ =\ 
\tilde\chi\,{_1\over^{[p]}}\,,
\label{kmvd}
\qqq
where $\,[p]$ stands for the $q$-deformation of $\,p\equiv2j+1$: 
$\,[p]\equiv{q^p-q^{-p} \over q-q^{-1}}$ with $q\equiv\ee^{-{\pi i\over
k+2}}$. The field $u$ will play the role of quantization 
of $\,h''_{_L}=h'_{_L}\m(\Pi-\Pi^{-1})^{-1}$, $\,\tilde u$ of 
quantization of ${h_{_L}'}^{-1}$ and $\tilde\chi$ as that 
of ${h_{_L}''}^{-1}$. On $\,\CH_{_L}$ we have the following commutation 
relations with the energy-momentum tensor 
\qq
[T(x)\m,\,\chi(x')]\ =\ -\m{_{3\pi\m i}\over^{2(k+2)}}\,\delta'(x-x')
\,\m\chi(x')\,+\,\,2\pi\m i\,\delta(x-x')\,\m\da\chi(x')
\label{wemt}
\qqq
and similarly for $\tilde\chi$, $u$ and $\tilde u$. They mean that 
all these fields are primary with conformal weight $\,\Delta_{1/2}
={3k\over4(k+2)}$
for the Virasoro action induced by $T$ on the cohomology of the Fock spaces.
We also have the commutation relations with the current
\qq
[J(x)_{_1}\m,\,\chi(x')_{_2}]\ =\ 2\pi\,\delta(x-x')\,C_{_{12}}\,
\chi(x')_{_{2}}
\label{wcur}
\qqq
and the same for $u$ and 
\qq
[J(x)_{_1}\m,\,\tilde\chi(x')_{_2}]\ =\ -2\pi\,\delta(x-x')\,
\tilde\chi(x')_{_{2}}\,C_{_{12}}\,
\label{wcut}
\qqq
and the same for $\tilde u$, all for $\vert x-x'\vert<2\pi$. They express 
the fact that the fields form primary spin $\hf$ multiplets of the current
algebra. Finally, we also have on $\CH_{_L}$ the exchange relations
\qq
\chi(x)_{_1}\,\chi(x')_{_2}&=&\chi(x')_{_2}\,\chi(x)_{_1}
\,\, D^\pm\,,\qquad\quad\tilde\chi(x)_{_1}\,\tilde\chi(x')_{_2}\ =\ 
\tilde D^{\pm}\,\ \tilde\chi(x')_{_2}\,\tilde\chi(x)_{_1}\,,
\label{chir}\\
\cr
u(x)_{_1}\,u(x')_{_2}&=& 
u(x')_{_2}\,u(x)_{_1}\,\,\tilde D^{'\pm}\,,\qquad\quad
\tilde u(x)_{_1}\,\tilde u(x')_{_2}\ =\ 
{D'}^{\pm}\,\,\tilde u(x')_{_2}\,\tilde u(x)_{_1}\,,
\label{ur}
\qqq
for ${_{x>x'}\atop^{x<x'}}$ and $|x-x'|<2\pi$. 
They may be established with the help of \,(\ref{comr}) and (\ref{bgcr}).
The matrix
\qq
D^\pm\ =\ q^{\mp\hf}\left(\matrix{1&0&0&0\cr0&q^{\pm1}{{[p+1]}
\over{[p]}}&q^{\mp(p-1)}{1\over{[p]}}&0\cr
0&-q^{\pm(p+1)}{1\over{[p]}}&q^{\pm1}{{[p-1]}\over{[p]}}&0\cr
0&0&0&1}\right)
\label{Dpd}
\qqq
represents a special $6j$-symbol\footnote{The matrix of $D$ corresponds 
to the lexicographic order of the basis vectors in the tensor product 
space.} dependent on $k$ and $j$ through $\m q=\ee^{-{\pi\m i\over k+2}}$ 
and $p=2j+1$. The matrix $\tilde D^\pm$ differs from $D^\pm$ by 
the interchange of the $12,12$ and $21,21$ elements on the diagonal.
The primed matrices $D^{'\pm}(p)$ and $\tilde D^{'\pm}(p)$ are equal
to the unprimed ones except for $p=k+1$ where, respectively, 
the third (second) entry on the diagonal is set to zero and for $p=1$ 
where the second (third) entry on the diagonal is put to zero.
\vskip 0.3cm

Let us explain how the latter modifications arise. The relation
\qq
\chi_{_{a1}}(x)\,\chi_{_{b2}}(x')\ =\ 
\chi_{_{b2}}(x')\,\chi_{_{a1}}(x)\,D^\pm_{_{1212}}
\ +\ \chi_{_{b1}}(x')\,\chi_{_{a2}}\,D^\pm_{_{2112}}\,,
\label{1212}
\qqq
becomes in the action on $\CV_{_{k,j}}$ with $j=k+1$ the identity
\qq
\chi_{_{a1}}(x)\,\chi_{_{b2}}(x')\ =\ 
\ =\ \chi_{_{b2}}(x')\,\chi_{_{a1}}(x)\,{_{q^{\pm\hf}\,
[k+2]}\over^{[k+1]}}\,-\,\chi_{_{b1}}(x')\,\chi_{_{a2}}
(x)\,{_{q^{\pm(k+{_3\over^2})}}\over^{[k+1]}}
\qqq
which is consistent with vanishing of the raising components
$\chi_{_{a2}}(x),\ \chi_{_{b2}}(x')$ since $[k+2]=0$. To obtain 
the commutation relations for the $u$'s, we have to divide both 
sides of (\ref{1212}) by $[p+1][p]$. \,This results in the relation
\qq
u_{_{a1}}(x)\,u_{_{b2}}(x')\ =\ 
u_{_{b2}}(x')\,u_{_{a1}}(x)\,\tilde D^\pm_{_{1212}}
\ +\ u_{_{b1}}(x')\,u_{_{a2}}(x)\,\tilde D^{\pm}_{_{2112}}\,.
\label{12122}
\qqq
for $p<k+1$. For $p=k+1$, however, the division by $[k+2]=0$
is not allowed and we have to replace $\tilde D^\pm_{_{1212}}(k+1)
=q^{\pm\hf}[2]$ by zero to obtain a true relation. The modifications
in the exchange relations of $\,\tilde u$'s follow similarly as do
the ones for $\,p=1$. In fact the commutation relations 
on $\,\CV_{_{k,j'}}\,$ for $\,j'={_k\over^2}-j\,$ may be obtained 
from those on $\,\CV_{_{k,j}}$ by simply interchanging the indices 
$\,1\leftrightarrow 2$.
\vskip 0.6cm

\subsection{Quantization of the boundary degrees of freedom}
\vskip 0.2cm

On the quantum group side, following \cite{FG}, one may realize 
the representations $\,\CV_{q,j}$ of $\,\CU_q(su(2))\,$ as a cohomology 
of modules carrying a representation of an algebra of deformed creators
and annihilators $\Pi^\pm,\,\Psi^\pm,\,B,\,\Gamma$ satisfying 
the relations:
\qq
&\Psi\Pi=q\Pi\Psi\,,\qquad\Pi B=B\Pi\,,\qquad\Pi\Gamma=\Gamma\Pi\,,&\cr
\label{dccr}\\
&q\,B\Gamma-q^{-1}\Gamma B=q-q^{-1}\,,\quad
\Psi B=qB\Psi\,,\quad\Psi\Gamma=q^{-1}\Gamma\Psi\,.&
\nonumber
\qqq
The latter may be represented in the space 
\qq
\CV\ \ =\mathop{\oplus}
\limits_{p\,{\rm mod}\,2(k+2)}\CV_{_{z\m q^p}}\,,
\label{cv}
\qqq
for $z$ a complex non-zero number, where $\,\CV_{_{z\m q^p}}$  
is spanned by the vectors $\,\vert\sigma,p\rangle\,$ with $\,\sigma=0,1,
\dots,k+1,$ by setting
\qq
&\Pi\,\vert\sigma,p\rangle=z\m q^p\vert\sigma,p\rangle\,,\quad
\Psi\,\vert\sigma,p\rangle=q^{-\sigma}\vert\sigma,\,p-1\,{\rm mod}\,2(k+2)
\rangle\,,&\cr
\label{real}\\
&B\,\vert\sigma,p\rangle=(1-q^{-2\sigma})\,\vert\sigma-1,
p\rangle\,,\quad\Gamma\,\vert\sigma,p\rangle=(1-\delta_{_{\sigma,p-1}})\,
\vert\sigma+1,p\rangle\,.&
\nonumber
\qqq
Recall that $\,\CU_q(su(2))\,$ is generated by $q^{\pm H}$, $E$ and $F$ 
with the relations
\qq
q^{\m H}\m E\,=\,E\,q^{\m H+2}\,,\quad q^{\m H}\m F\,=\,F\,q^{\m H-2}\,,
\quad[E\m,\,F]\,=\,{{q^{\m H}-q^{-H}}\over{q-q^{-1}}}\,,
\label{qg}
\qqq
which are a deformation of the standard relations for $su(2)$.
The spin $j=0,1,\dots,{k+1\over2}$ irreducible highest weight 
modules $\,\CV_{_{q,j}}$ of $\,\CU_q(su(2))\,$ have dimension 
$2j+1$ and are generated from an eigen-vector of $\,q^{\pm H}\,$ 
with eigenvalue $\,q^{\pm 2j}$ annihilated by $E$ and by $F^{2j+1}$,
the highest weight vector. \,Upon introduction of the $2\times2$ 
matrices built of the $\,\CU_q(su(2))\,$ generators
\qq
\gamma\ \equiv\ \left(\matrix{q^{\m H}&\ q\m(q-q^{-1})\m E\cr
(q-q^{-1})\m F\,q^H&\ q(q-q^{-1})^2\m F\m E+q^{-H}}\right)\,,
\label{qg0}
\qqq
the commutation relation (\ref{qg}) may be rewritten 
in the matrix form
\qq
\gamma_{_1}\m R^+\gamma_{_2}(R^\mp)^{-1}\ =\ R^\pm\gamma_{_2}(R^-)^{-1}
\gamma_{_1}\,,
\label{qg1}
\qqq
which is more convenient for algebraic manipulations.
The $4\times4$ $\,R$-matrices are given by
\qq
R^+\ =\ q^{\m\hf}\left(\matrix{q^{-1}&0&0&0\cr0&1&q^{-1}-q&0\cr
0&0&1&0\cr0&0&0&q^{-1}}\right),\qquad
R^-\ =\ q^{-\hf}\left(\matrix{q&0&0&0\cr0&1&0&0\cr
0&q-q^{-1}&1&0\cr0&0&0&q}\right)\,.\hspace{0.3cm}
\qqq
Note a useful property: $\,(R^\pm)^{-1}=P\m R^\mp P$, 
\,where $P$ exchanges the factors in the tensor product.
\vskip 0.3cm

One may turn the spaces $\CV_{_{z\m q^p}}$ into $\,\CU_q(su(2))$-modules 
by expressing the matrices $\,\gamma\,$ by the deformed creators
and annihilators:
\qq
\gamma\ =\ q\left(\matrix{\Pi(1-B\Gamma)&\ -(\Pi-\Pi^{-1})B+\Pi B\Gamma B\cr
-\Pi\Gamma&\ \Pi^{-1}+\Pi\Gamma B}\right).
\label{qg01}
\qqq
The commutation relations (\ref{qg}) follow from (\ref{dccr}). 
For $z$ not an (integer) power of $q$, 
the $\,\CU_q(su(2))$-modules $\,\CV_{_{z\m q^p}}$ are irreducible. 
This is in general not the case for $z$ a power of $q$. To obtain 
irreducible modules for $z=1$ (it is clearly enough to consider this case), 
one defines on $\,\CV\,$ the screening operator 
$\,\CQ=\Pi\m\Gamma\Psi^{-2}$ which maps $\CV_{_{q^p}}$ to $\CV_{_{q^{p+2}}}$ 
and is nilpotent: $\,\CQ^{k+2}=0$. It gives rise \cite{FG} to the complexes
\qq
0\ \longrightarrow\ \CV_{_{q^{-p}}}\ \mathop{\longrightarrow}\limits^{\CQ^p}
\ \ \CV_{_{q^p}}\ \mathop{\longrightarrow}\limits^{\CQ^{k+2-p}}\ 
\CV_{_{q^{-p}}}\ \longrightarrow\ 0\,,\cr\cr
0\ \longrightarrow\ \,\CV_{_{q^p}}\,\ \mathop{\longrightarrow}
\limits^{\CQ^{k+2-p}}\ \CV_{_{q^{-p}}}\ \mathop{\longrightarrow}
\limits^{\CQ^{p}}\ \,\CV_{_{q^{p}}}\,\ \longrightarrow\ 0\,\ \,
\qqq
of  $\,\CU_q(su(2))$-modules exact in the middle.
It follows that the action of $\,\CU_q(su(2))$ descends to the
cohomology spaces
\qq
\tilde\CV_{_{q^p}}\ =\ \cases{\CV_{_{q^p}}\m/\m\CQ^p(\CV_{_{q^{-p}}})
\qquad{\rm for}\ \ p=0,1,\dots,k+2\,,\cr
{\rm ker}\,\CQ^{-p}\subset\CV_{_{q^p}}\qquad{\rm for}\ \ p=0,-1,
\dots,-k-2\,.}
\label{qsp}
\qqq
Note that $\tilde\CV_{_1}=\{0\}$ and that $\CQ^{k+2-p}$ induces
an isomorphism of $\tilde\CV_{_{q^p}}$ and $\tilde\CV_{_{q^{-p}}}$. 
For $j=0,{1\over2},\dots,{k+1\over^2}$ and $p=2j+1$, the 
$\,\CU_q(su(2))$-modules $\,\tilde\CV_{_{q^p}}$ may be identified
with the irreducible highest weight modules $\,\CV_{_{q,j}}$ of 
$\m\,\CU_q(su(2))$ of spin $j$ with the highest weight 
vector corresponding to  $\vert0,p\rangle\in\CV_{_{q^p}}$. The matrix 
$\gamma$ of generators may be decomposed in the action on 
$\,\tilde\CV_{_{q^p}}$ into the upper- and lower-triangular 
parts\footnote{This requires
a choice of the square roots $q^{\pm{H\over 2}}$ of $q^{\pm H}$ which
may be fixed by demanding that $q^{\pm{H\over 2}}\vert0,p\rangle
=q^j\vert0,2j+1\rangle$ but is irrelevant.} $\gamma_{_\pm}$
such that $\,\gamma=\gamma_{_-}\gamma_{_+}^{-1}$ with
\qq
\gamma_{_+}\,=\left(\matrix{q^{-{H\over 2}}&\ -(q-q^{-1})\m q^{-{H\over 2}}E
\cr0&\ \ q^{\m{H\over 2}}}\right),\quad
\gamma_{_-}\,=\left(\matrix{q^{\m{H\over 2}}&\ 0\cr(q-q^{-1})\m F\,
q^{\m{H\over 2}}&\ q^{-{H\over 2}}}\right)
\qqq
In terms of $\gamma_{_\pm}$, the commutation relations (\ref{qg1})
become
\qq
&&{\gamma_{_+}}_{_1}{\gamma_{_+}}_{_2}\,R^\pm
\ =\ R^\pm\,{\gamma_{_+}}_{_2}{\gamma_{_+}}_{_1}\,,\ \quad
{\gamma_{_-}}_{_1}{\gamma_{_-}}_{_2}\,R^\pm
\ =\ R^\pm\,{\gamma_{_-}}_{_2}{\gamma_{_-}}_{_1}\,,\cr\label{gpm}\\
&&{\gamma_{_+}}_{_1}{\gamma_{_-}}_{_2}\, R^+
\ =\ R^+\,{\gamma_{_-}}_{_2}{\gamma_{_+}}_{_1}\,,\ \quad
{\gamma_{_-}}_{_1}{\gamma_{_+}}_{_2}\,R^-\ =\ 
R^-\,{\gamma_{_+}}_{_2}{\gamma_{_-}}_{_1}\,.
\qqq
\vskip 0.3cm

In order to quantize the monodromy data $\,U$, see (\ref{us}),
we shall need also the ``quantum group vertex operators'', see
\cite{FG,FHT}. Let us introduce two matrices of operators acting in 
the space $\,\CV\m$:
\qq
a\,&=&\left(\matrix{-\Psi&qB\Psi\cr
\Psi\CQ&q(\Pi-\Pi^{-1})\Psi^{-1}
-q B\Psi\CQ}\right){1\over{q-q^{-1}}},\cr\cr\cr
\tilde a\,&=&\left(\matrix {-(\Pi-\Pi^{-1})\Psi^{-1}+B\Psi\CQ&B\Psi\cr
\Psi\CQ&\Psi}\right).\hspace{0.4cm}
\label{qgvo}
\qqq
They satisfy the relations
\qq
\tilde a\m\,a\ =\ {_{\Pi-\Pi^{-1}}\over^{q-q^{-1}}}\,,\qquad
\gamma_{_1}\,a_{_2}\ =\ a_{_2}\m(R^-)^{-1}\gamma_{_1}\m R^+\,,
\qquad\tilde a_{_1}\,\gamma_{_2}\ =\ R^+\m\gamma_{_2}\,(R^-)^{-1}
\,\tilde a_{_1}\,.
\label{aga}
\qqq
The components
$\,a_{{1a}}$ and $\,\tilde a_{{a2}}$ map $\,\CV_{_{z\m q^p}}$ into 
$\,\CV_{_{z\m q^{p-1}}}$ and, for $z=1$, pass to the quotient spaces 
(\ref{qsp}):
\qq
a_{{1a}}:\ \tilde\CV_{_{q^p}}\ \longrightarrow\ \tilde\CV_{_{q^{p-1}}}\,,
\qquad\tilde a_{{a2}}:\ \tilde\CV_{_{q^p}}\ \longrightarrow\ 
\tilde\CV_{_{q^{p-1}}}\,.
\qqq
Similarly, the components $\,a_{{2a}}$ and $\,\tilde a_{{a1}}$ map 
$\,\CV_{_{z\m q^p}}$ into $\,\CV_{_{z\m q^{p+1}}}$ and pass to the quotients:
\qq
a_{{2a}}:\ \tilde\CV_{_{q^p}}\ \longrightarrow\ 
\tilde\CV_{_{q^{p+1}}}\,,\qquad\tilde a_{{a1}}:\ \tilde\CV_{_{q^p}}
\ \longrightarrow\ \tilde\CV_{_{q^{p+1}}}\,.
\label{116}
\qqq
We shall also need modified version of the vertex operators
defined by
\qq
b\ =\ a\,{_{q-q^{-1}}\over^{\Pi-\Pi^{-1}}}
\label{b}
\qqq
for generic values of $z$. $\,b\m$ is the inverse\footnote{In
\cite{FG}, $\tilde a$ was denoted  by $g_0$ and $\,b$ by $g_0^{-1}$; 
the lower indices 1 and 2 of $\,a\,$ in (\ref{116})
and (\ref{b}) correspond to the upper indices 2 and 1 in \cite{FHT}.}
of the matrix $\tilde a\m$: $\,b\,\tilde a=\tilde a\,b=1$.
For $z=1$, $\,b$ may be still defined on the components  
$\,\CV_{_{q^p}}$ and $\,\tilde \CV_{_{q^p}}$ such that $[p]\not=0$.
The components of $a$, $\tilde a$ and $b$ satisfy
for generic $z$ the commutation relations which may be
summarized as
\qq
a_{_1}\m\,a_{_2}=(\tilde D^{\pm})^{-1}\,a_{_2}\m\,a_{_1}\, R^\pm\,,
\quad
\tilde a_{_1}\m\,\tilde a_{_2}=R^\pm\,\tilde a_{_2}\m\,\tilde a_{_1}
\,(D^\pm)^{-1}\,,\quad b_{_1}\m\,b_{_2}=(D^{\pm})^{-1}\,b_{_2}\m\,
b_{_1}\, R^\pm
\qquad
\label {ata}
\qqq
with the same $p$-dependent matrices $D^\pm$ and $\tilde D^\pm$
as introduced in the previous section, see (\ref{Dpd}), except for
the replacement of $q^{\pm p}$ by the eigenvalues of $\Pi^\pm$
equal to $z^\pm q^{\pm p}$.  For $z=1$ the first two relations
of (\ref{ata}) still hold whenever they do not involve division
by a vanishing $[p]$, i.e.\,\,away from the boundary values $\,p=0,k+2$,
whereas the $\m3^{\m\rm rd}$ relation requires corrections
(as the for the the Kac-Moody fields $\m u,\m\tilde u$).
\vskip 0.3cm

Let us fix $z=1$ and the spins $j_{_0}$ and $j_{_\pi}$ between $0$ 
and $k\over2$ and define the matrix of monodromy operators
\qq
\tilde U\ =\ a\,\,\gamma_{_{\pi-}}^{-1}\m\gamma_{_{\pi+}}\,\tilde a\,.
\label{tu}
\qqq
acting on the space
\qq
\tilde\CV\ \ =\mathop{\oplus}\limits_
{p\,\m{\rm mod}\m\,2(k+2)}\tilde\CV_{_{q^{p_{_0}}}}
\otimes\tilde\CV_{_{q^{p_{_\pi}}}}\otimes\tilde\CV_{_{q^p}}
\label{ds}
\qqq
where $p_{_0}=2j_{_0}+1$ and $p_{_\pi}=2j_{_\pi}+1$, $\,\gamma_{_{\pi\pm}}$
acts on the $\,\tilde\CV_{_{q^{p_{_\pi}}}}$ factor and $\,a\,$ and 
$\,\tilde a\,$ on $\,\tilde\CV_{_{q^p}}$. $\,\tilde U$ quantizes
the classical monodromy matrix $\,U$, see (\ref{us}), modulo potentially 
singular factors involving $(\Pi-\Pi)^{-1}$ that we have redistributed 
to the Kac-Moody vertex operators. Note that $\tilde U_{11}$ and 
$\tilde U_{22}$ preserve the value of $p$ labeling the $3^{\m\rm rd}$ 
representation space $\m\tilde\CV_{_{q^p}}$, $\m\tilde U_{12}$ lowers 
it by $2$ and $\tilde U_{21}$ raises it by $2$ (mod $2(k+2)$). Recall 
that $\,\tilde\CV_{_{q^p}}$ carries the spin $j$ irreducible representation 
of $\,\CU_q(su(2))$ for $\vert p\vert=2j+1$ so that the direct sum 
in (\ref{ds}) contains each spin $j$ factor between $0$ and ${k\over2}$ 
twice and spin ${k+1\over 2}$ once. 
\vskip 0.3cm

The tensor products $\,\tilde\CV_{_{q^{p_{_0}}}}
\otimes\tilde\CV_{_{q^{p_{_\pi}}}}\otimes\tilde\CV_{_{q^p}}$ carry 
the co-product action of $\,\CU_q(su(2))$ defined so that the matrix $\gamma$
of generators of $\,\CU_q(su(2))$, see (\ref{qg}), acts as
$\,\gamma_{_{0-}}\gamma_{_{\pi-}}\gamma_{_w}
\,\gamma_{_{\pi+}}^{-1}\gamma_{_{0+}}^{-1}\equiv\Delta\gamma$, 
\,where $\gamma_{_w}$ represents the $\,\CU_q(su(2))$-action in the 
$3^{\m\rm rd}$ factor $\,\tilde\CV_{_{q^p}}$. An important property
of $\,\tilde U$, proven in Appendix 1, is that it commutes with
the co-product action of $\,\CU_q(su(2))$. It follows then that 
$\,\tilde U\,$ preserves the subspace of the invariant
tensors of $\,\tilde\CV$,
\qq
\tilde\CV^{^{inv}}\ =\ \mathop{\oplus}\limits_
{p\,\m{\rm mod}\m\,2(k+2)}(\tilde\CV_{_{q^{p_{_0}}}}
\otimes\tilde\CV_{_{q^{p_{_\pi}}}}\otimes\tilde\CV_{_{q^p}})^{^{inv}}
\qqq
i.e.\,\,tensors in the kernel of $\,(\Delta\gamma\m-\m 1)\m$ where $1$ 
denotes the $2\times2$ unit matrix. For $0<j_{_0},\m j_{_\pi},\m 
j<{k+1\over2}$, the dimensions of the spaces of invariant tensors 
are as in the undeformed case \cite{FHT}, i.e.\,\,they are given by 
the standard angular momentum rule:
\qq
(\tilde\CV_{_{q^{p_{_0}}}}
\otimes\tilde\CV_{_{q^{p_{_\pi}}}}\otimes\tilde\CV_{_{q^p}})^{^{inv}}
\cong\,\cases{\,\NC\ \,{\rm if}\ \,\vert j_{_0}-j_{_\pi}\vert\,\leq\,
j\,\leq\,j_{_0}+j_{_\pi},\ \,j_{_0}+j_{_\pi}+j=0\,\m{\rm mod}\m\,1\,,\cr
\{0\}\ {\rm otherwise}.}
\nonumber
\qqq
The action of $\,\tilde U\m$ on the spaces of invariants may be found 
by an explicit computation. One obtains for positive $p$ and the diagonal 
matrix elements of $\,\tilde U$,
\qq
&&\tilde U_{11}\ =\ -\m{q^{-1}(q^{\m p_{_0}}+q^{-p_{_0}})-q^p(q^{\m 
p_{_\pi}}+q^{-p_{_\pi}})\over q-q^{-1}}\,,\label{dme}\\\cr
&&\tilde U_{22}\ =\ \,\,{q^{-1}(q^{\m p_{_0}}+q^{-p_{_0}})-q^{-p}
(q^{\m p_{_\pi}}+q^{-p_{_\pi}})\over q-q^{-1}}\,.\label{dmep}
\qqq
The non-diagonal matrix elements of $\,\tilde U\,$ are for a special 
choice of the basis of invariant tensors given by:
\qq
&&\tilde U_{12}\ =\ q^{-1}(q-q^{-1})\,[j_{_0}+j_{_\pi}+j+1]\,[j_{_\pi}
+j-j_{_0}]\,[j_{_0}+j-j_{_\pi}]\,,\label{ndme}\\\cr
&&\tilde U_{21}\ =\ q^{-1}(q-q^{-1})\,[j_{_0}+j_{_\pi}-j]\,.\label{ndmep}
\qqq
The value $\,p=2j+1\,$ refers to the tensor in the domain and
is lowered by 2 by $\,\tilde U_{12}$ and raised by $\,\tilde U_{21}$. 
\,The diagonal combinations that are basis independent have the form
\qq
&&\tilde U_{12}\,\tilde U_{21}\,=\,q^{-2}(q-q^{-1})^2\,[j_{_0}+j_{_\pi}+j+2]
\,[j_{_\pi}+j-j_0+1]\,[j_{_0}+j-j_{_\pi}+1]\,[j_{_0}+j_{_\pi}-j]\,,\cr\cr
&&\tilde U_{21}\,\tilde U_{12}\,=\,q^{-2}(q-q^{-1})^2\,[j_{_0}+j_{_\pi}-j+1]
\,[j_{_0}+j_{_\pi}+j+1]\,[j_{_\pi}+j-j_0]\,[j_{_0}+j-j_{_\pi}]\,.
\nonumber
\qqq
The space $\,(\tilde\CV_{_{q^{p_{_0}}}}\otimes\tilde\CV_{_{q^{p_{_\pi}}}}
\otimes\tilde\CV_{_{q^p}})^{^{inv}}\,$ is composed of null-vectors 
of the natural hermitian form if and only if $p_{_0}$, $p_{_\pi}$
and $p$ violate the fusion rule $\,j_{_0}+j_{_\pi}+j\leq k$.
Note that $\,\tilde U_{12}\m\tilde U_{21}$ vanishes
when applied to $\,(\tilde\CV_{_{q^{p_{_0}}}}\otimes\tilde\CV_{_{q^{p_{_\pi}}}}
\otimes\tilde\CV_{_{q^p}})^{^{inv}}\,$ with $\,j_{_0}+j_{_\pi}+j=k\,$
due to the vanishing of $\,[j_{_0}+j_{_\pi}+j+2]=[k+2]$
and, similarly, $\,\tilde U_{21}\m\tilde U_{12}$ vanishes
for $\,j_{_0}+j_{_\pi}+j=k+1$. \,In fact, it is the lowering operator
$\tilde U_{12}$ that annihilates $\,(\tilde\CV_{_{q^{p_{_0}}}}\otimes
\tilde\CV_{_{q^{p_{_\pi}}}}\otimes\tilde\CV_{_{q^p}})^{^{inv}}\,$ 
with $\,j_{_0}+j_{_\pi}+j=k+1\,$.
The relations for the negative $p$ are a mirror image of those for
the positive $p$ so that, for negative $p$, it is the raising operator 
$\,\tilde U_{21}$ that vanishes in the action on 
$\,(\tilde\CV_{_{q^{p_{_0}}}}\otimes\tilde\CV_{_{q^{p_{_\pi}}}}
\otimes\tilde\CV_{_{q^p}})^{^{inv}}\,$ with $\,j_{_0}+j_{_\pi}+j=k+1\,$.
These are the only matrix elements that could lead out of the
subspace of the null invariant tensors. It follows that
$\,\tilde U_{ij}$ preserve this subspace and, hence,  descend
to the quotient space on which the hermitian form defines
a positive scalar product. The quotient space may be identified with
\qq
\mathop{\oplus\m'}\limits_{j\ }\m\,\,
(\tilde\CV_{_{q^{p_{_0}}}}\otimes\tilde\CV_{_{q^{p_{_\pi}}}}\otimes
\tilde\CV_{_{q^p}})^{^{inv}}\,
\qqq
with $\oplus\m'$ denoting the sum restricted to $j$
such that $\,\vert j_{_0}-j_{_\pi}\vert\,\leq\,j\,\leq\,{\rm min}
(j_{_0}+j_{_\pi},\m\,k-j_{_0}-j_{_\pi})\,$ and $\,j_{_0}+j_{_\pi}+j=0
\,\m{\rm mod}\m\,1$. $\,$Note that $j={k+1\over2}$ is necessarily absent 
in the sum which falls into two isomorphic terms involving, respectively, 
the positive and the negative $p$ with no non-zero matrix elements 
of $\,\tilde U$ between them. We shall identify the term with
positive $p$ with the boundary space of states, see the end 
of Sect.\,\,6.1\m:
\qq
\CH^{\,bd}_{_{j_{_0}j_{_\pi}}}\ =\ \mathop{\oplus\m''}\limits_{j\ \,}
\m\,\,(\tilde\CV_{_{q^{p_{_0}}}}\otimes\tilde\CV_{_{q^{p_{_\pi}}}}
\otimes\tilde\CV_{_{q^p}})^{^{inv}}\,,
\label{inv1}    
\qqq
where the sum $\m\oplus\m''$ is as $\oplus\m'$ but with $p$ restricted
to the positive values. 
\vskip 0.9cm

\nsection{Local bulk fields in the boundary WZW $SU(2)$ theory}
\subsection{Quantization of the classical bulk fields}
\vskip 0.2cm

According to the preceding discussion, see (\ref{ndebd}), the Hilbert space 
of the complete boundary theory, combining the bulk Kac-Moody degrees 
of freedom and the boundary quantum group ones is
\qq
\CH_{_{j_{_0}j_{_\pi}}}\ =\ \mathop{\oplus\m''}\limits_{j\ \,}\m\,\,
\CH_{_{j_{_0}j_{_\pi}j}}\ \,\subset\,\ \CH_{_L}\otimes\CH^{bd}_{_{j_{_0}
j_{_\pi}}}\,,
\label{cHs0}
\qqq
where
\qq
\CH_{_{j_{_0}j_{_\pi}j}}\ \equiv\ \CV_{_{k,j}}
\otimes\,(\tilde\CV_{_{q^{p_{_0}}}}\otimes\tilde
\CV_{_{q^{p_{_\pi}}}}\otimes\tilde\CV_{_{q^p}})^{^{inv}}
\label{cHs}
\qqq
with $p=2j+1$.  $\,\CH_{_L}$ is given by (\ref{hel}). We are now 
prepared for quantization of the classical bulk fields $g(t,x)$ defined
as functions on the phase space $\,\CP_{_{\mu_{_0}\mu_{_\pi}}}$
by (\ref{do0pi}). On the quantum level, we shall define 
\qq
g(t,x)\ =\ u(t+x-2\pi)\m\,\tilde U\,\,\tilde u(t-x)\,,
\label{qfg}
\qqq
see (\ref{kmvo}) and (\ref{kmvd}). Note that the matrix elements
of $\,g(t,x)$, \,acting {\it a priori\,} in 
$\,\CH_{_L}\otimes\CH^{bd}_{_{j_{_0}j_{_\pi}}}$, preserve 
the diagonal subspace $\,\CH_{_{j_{_0}j_{_\pi}}}$
and we shall consider them as operators on that space, i.e.\,\,on 
the Hilbert space of the boundary theory. Indeed, in components,
\qq
g_{ab}(t,x)\ =\ u_{ai}(t+x-2\pi)\m\,\tilde U_{i\ell}\,\tilde u_{\ell b}
(t-x)\,.
\label{cmp}
\qqq
But $u_{a1}\tilde u_{b1}$ and $u_{a1}\tilde u_{b1}$ preserve the spin 
$j$ of the  Kac-Moody representations, $u_{a1}\tilde u_{b2}$ lowers
it by 1 whereas $u_{a2}\tilde u_{b1}$ raises it by 1, just like
$\tilde U_{i\ell}$ do for the quantum group spin $j$. 
Definition (\ref{qfg}) is inspired by the classical formula 
(\ref{do0pi}). Quantum fields $\,g(t,x)\,$ satisfy the commutation relations 
that follow from (\ref{wcur}), (\ref{wcut}) and (\ref{wemt}) and
encode their current and conformal algebra symmetries:
\qq
[J(t+x)_{_1}\m,\,g(t,x')_{_2}]&=&2\pi\,\delta(x-x')\m\,C_{_{12}}\,
g(t,x')_{_{2}}\ -\ 2\pi\,\delta(x+x')\m\,g(t,x')_{_2}\,C_{_{12}}\,,\cr\cr
[T(t+x)\,,\m\,g(t,x')]\ \,\m&=&-\m{_{3\pi\m i}\over^{2(k+2)}}\,(\delta'(x-x')+
\delta'(x+x'))\,\m g(t,x')\cr
&&\hspace{1.85cm}+\,2\pi\m i\,(\delta(x-x')\,\da_{_+}
-\m\delta(x+x')\,\da_{_-})\,\m g(t,x')
\nonumber
\qqq
for $\vert x\vert\leq2\pi$.
The most important property of $g(t,x)$ is the locality:
\qq
[g(t,x)_{_1}\m,\, g(t,x')_{_2}]\ =\ 0\qquad {\rm for}\quad x\not=x'\,,
\label{loc}
\qqq
i.e.\,\,the commutativity at different spatial points, a basic property
of local quantum field theory considered here on the space time
with a finite spatial extension.
\vskip 0.6cm

\subsection{Locality of the bulk fields: generic case}
\vskip 0.2cm

We shall show the locality of $\,g(t,x)$, \m the main result of this
section, in two steps. First we shall demonstrate that (\ref{loc}) holds 
on the components of the Hilbert space (\ref{cHs0}) that do not involve 
extreme values of spins. This will be done by multiple application 
of the commutation rules between the building blocks
of field $\,g$. \,To perform this calculation,
it will be convenient to reshuffle again the ${1\over[p]}$ factors 
and to introduce a modified version of the monodromy matrix defined 
for generic values of $z$ labeling the representations of the 
$q$-deformed creator and annihilator algebra (\ref{dccr}). We set
\qq
U\ =\ b\,\,\gamma_{_{\pi-}}^{-1}\m\gamma_{_{\pi+}}\,\tilde a\,,
\label{tb}
\qqq
where $b$ is the matrix of 
the modified quantum group vertex operators given by (\ref{b}).
The components of $\,U\,$ act on the space $\CV$ of (\ref{cv}) 
with generic $z$. Using the same formula for $z=1$, one may also 
define the components $\,U_{i\ell}$ as operators on the subspace 
of $\,\tilde\CV$ of (\ref{ds}) with $\m[p+\delta_{1\ell}-\delta_{\ell2}]
\not=0\m$ and one has then the relation
\qq
U_{i\ell}\ =\ {_1\over^{[p+\delta_{i1}-\delta_{i2}]}}\,\tilde U_{i\ell}
\ =\ \tilde U_{i\ell}\,{_1\over^{[p+\delta_{\ell1}-\delta_{\ell2}]}}\,.
\qqq
to the components of the monodromy matrix $\tilde U$ of (\ref{tu}). 
We show in Appendix 2 the following commutation relation
\qq
U_{_1}\,(D^+)^{-1}\, U_{_2}\,D^\mp\ =\ (D^\pm)^{-1}\, U_{_2}\m\,D^-\,
U_{_1}
\label{ada}
\qqq
holding for generic $z$ and for $z=1$ whenever no division 
by vanishing $[p]$ is involved. Note a similarity with the
relation (\ref{qg1}) with matrices $\,(D^\pm)^{-1}$ replacing
$\,R^\pm$. 
\vskip 0.3cm

The quantum field (\ref{qfg}) may be rewritten with the use of the 
modified monodromy monodromy  $\m U\m$ as
\qq
g(t,x)\ =\ \chi(t+x-2\pi)\m\,U\,\,\tilde u(t-x)\,,
\label{qfh}
\qqq
whenever the right hand side is well defined, see (\ref{kmvd}).
The commutation relations (\ref{chir}) and (\ref{ur}) between 
the vertex operators $\chi$ and $\tilde u$ imply also that,
away from the subspaces with vanishing $[p]$,
\qq
\tilde u(x)_{_1}\m\,\chi(x')_{_2}\ =\ \chi(x')_{_2}\m\,(D^\pm)^{-1}\,
\tilde u(x)_{_1}\,.
\label{qrh}
\qqq
Employing this rule, we obtain
\qq
&&g(t,x)_{_1}\,\,g(t,x')_{_2}\ =\ \chi(t+x-2\pi)_{_1}\m\,U_{_1}\m\,
\tilde u(t-x)_{_1}\m\,\chi(t+x'-2\pi)_{_2}\m\,U_{_2}\m\,
\tilde u(t-x')_{_2}\cr\cr
&&=\ \chi(t+x-2\pi)_{_1}\m\,U_{_1}\m\,
\chi(t+x'-2\pi)_{_2}\m\,(D^+)^{-1}\m\,\tilde u(t-x)_{_1}\m\,
U_{_2}\m\,\tilde u(t-x')_{_2} 
\qqq
or, in components,
\qq
g_{ab}(t,x)\,\,g_{cd}(t,x')\ =\ \chi_{ai}(t+x-2\pi)\,\m\,U_{i\ell}\,\,
\chi_{cn}(t+x'-2\pi)\hspace{0.8cm}\cr\cr
\cdot\,(D^+(\hat p))^{^{-1}}_{\m\,\ell n,mr}\,\,
\tilde u_{mb}(t-x)\,\,U_{rs}\,\,\tilde u_{sd}(t-x')\,.
\qqq
Above, $\hat p=2\hat j+1$ with $\hat j$ equal to the value of spin of
the Kac-Moody representation and we reserve the notation $p=2j+1$ 
for the labels of the quantum group representations. In general,
the latter are equal to the Kac-Moody ones only in the initial 
and final states.
Taking into account the shifts of the spins, we may rewrite 
the last relation as
\qq
&&g_{ab}(t,x)\,\,g_{cd}(t,x')\ =\ \chi_{ai}(t+x-2\pi)\,\m\,
\chi_{cn}(t+x'-2\pi)\cr\cr
&&\cdot\ U_{i\ell}\,\m\,(D^+(p+\delta_{1m}-\delta_{2m}+\delta_{1r}
-\delta_{2r}))^{^{-1}}_{\m\,\ell n,mr}\,\,U_{rs}\,\,\tilde u_{mb}(t-x)
\,\,\tilde u_{sd}(t-x')\,.
\qqq
But direct inspection shows that $\,D^+(p+\delta_{1m}-\delta_{2m}
+\delta_{1r}-\delta_{2r})=D^+(p)\,$ (the ``ice property'' of the $6j$ 
symbols) and we may use the relation (\ref{ada}) in order to obtain 
\qq
g(t,x)_{_1}\,\,g(t,x')_{_2}\ =\ \chi(t+x-2\pi)_{_1}\m\,\chi(t+x'-2\pi)_{_2}
\,\,(D^\pm(p))^{^{-1}}\m\,U_{_2}\,\cr\cr
\cdot\,D^-(p)\,\,U_{_1}\,\,(D^\mp(p))^{^{-1}}\,\,\tilde u
(t-x)_{_1}\m\,\tilde u(t-x')_{_2}\,.
\qqq
Again the extreme $\,D^\pm(p)\,$ may be replaced by $\,D^\pm(\hat p)\,$
which permits to use the commutation relations (\ref{chir})
and (\ref{ur}) to infer that
\qq
&g(t,x)_{_1}\,\,g(t,x')_{_2}\ =\ \chi(t+x'-2\pi)_{_2}\m\,\chi(t+x-2\pi)_{_1}
\,\,U_{_2}\m\,D^-(p)\,\,U_{_1}\,\,\tilde u(t-x')_{_2}\m\,
\tilde u(t-x)_{_1}&\cr\cr
&\ =\ \chi(t+x'-2\pi)_{_2}\,\,U_{_2}\m\,\chi(t+x-2\pi)_{_1}
\m\,D^-(p)\,\,\tilde u(t-x')_{_2}\m\,U_{_1}\,\,\tilde u(t-x)_{_1}\,.&
\qqq
Changing back $D^-(p)$ to $D^-(\hat p)$ and applying the commutation 
relation $\,\chi(x)_{_1}\m D^\pm\m\tilde u(x')_{_2}=\tilde u(x')_{_2}\m
\chi(x)_{_1}\,$ equivalent to (\ref{qrh}), we obtain in turn
\qq
g(t,x)_{_1}\,\,g(t,x')_{_2}\ =\ \chi(t+x'-2\pi)_{_2}\m\,U_{_2}\m\,
\tilde u(t-x')_{_2}\,\,\chi(t+x-2\pi)_{_1}\m\,U_{_1}\m\,
\tilde u(t-x)_{_1}\cr\cr
=\ g(t,x')_{_2}\,\,g(t,x)_{_1}\hspace{2.5cm}
\qqq
which settles the locality issue for the matrix elements of the
field $\,g(t,x)\,$ which do not involve the extreme values of spin.
\vskip 0.6cm

\subsection{Locality of the bulk fields: completion of the proof}
\vskip 0.2cm

We still have to look more closely at the commutation relations
on the components (\ref{cHs0}) of the physical Hilbert space 
close to the extreme values. This will involve a somewhat tedious
case by case inspection. Let us rewrite $g(t,x)$ explicitly 
in terms of the components raising by 1, lowering by 1 and preserving 
the value of spin $\m j\m$:
\qq
g_{ab}(t,x)\ =\ g^+_{ab}(t,x)\,+\,g^-_{ab}(t,x)\,+\,g^0_{ab}(t,x)\,,
\label{lrp}
\qqq
where
\qq
&&g^+_{ab}(t,x)=u_{a2}(t+x-2\pi)\m\,\tilde U_{21}\,\tilde u_{1b}(t-x)
\,,\quad
g^-_{ab}(t,x)=u_{a1}(t+x-2\pi)\m\,\tilde U_{12}\,\tilde u_{2b}(t-x)
\,,\cr
&&g^0_{ab}(t,x)=u_{a1}(t+x-2\pi)\m\,\tilde U_{11}\,\tilde u_{1b}(t-x)
\,+\,u_{a2}(t+x-2\pi)\m\,\tilde U_{22}\,\tilde u_{2b}(t-x)\,.
\nonumber
\qqq
The locality statement (\ref{loc}) is equivalent to the equations
\qq
&&[g^+_{ab}(t,x)\m,\,g^+_{cd}(t,x')]\ =\ 0\ =\ [g^-_{ab}(t,x)\m,\,g^-_{cd}
(t,x')]\,,\label{1}\\
&&[g^+_{ab}(t,x)\m,\,g^0_{cd}(t,x')]\,+\,[g^0_{ab}(t,x)\m,\,g^+_{cd}(t,x')]
\ =\ 0\,,\label{2}\\
&&[g^-_{ab}(t,x)\m,\,g^0_{cd}(t,x')]\,+\,[g^0_{ab}(t,x)\m,\,g^-_{cd}(t,x')]
\ =\ 0\,,\label{3}\\
&&[g^+_{ab}(t,x)\m,\,g^-_{cd}(t,x')]\,+\,[g^-_{ab}(t,x)\m,\,g^+_{cd}(t,x')]
\,+\,[g^0_{ab}(t,x)\m,\,g^0_{cd}(t,x')]\ =\ 0\,.\label{4}
\qqq
We still have to verify the above relations in the action on
the spin $j={k-1\over 2},{k\over2}$ and spin $j=0,{1\over2}$ 
components $\,\CH_{_{j_{_0}j_{_\pi}j}}$ of the Hilbert space  
(\ref{cHs}).
\vskip 0.3cm

Consider first the case with $j={k-1\over2}$. In this instant,
only the $\,[g^+,\m g^+]\,$ commutation relation of (\ref{1}) does
not follow from the previous considerations but is assured
anyway since the combination $g^+g^+$ vanishes on
$\,\CH_{_{j_{_0}j_{_\pi}{_{k-1}\over^2}}}$. The next case
$j={k\over2}$ will require, however, some work. 
Now the $\,[g^-,\m g^-]=0\,$ relation of (\ref{1}) follows from 
the previous considerations and the $\,[g^+,\m g^+]=0\,$ one and (\ref{2}) 
result from the vanishing on  $\,\CH_{_{j_{_0}j_{_\pi}{_k\over^2}}}$ 
of the raising components $\,\tilde u_{1a}$ of the Kac-Moody vertex 
operator implying that $g^+=0$ on that space. As for the other commutators,
\qq
&&\quad[g_{ab}^+(t,x),\m g_{cd}^-(t,x')]\cr\cr
&&=\ \Big(u_{a2}(t+x-2\pi)\m\,\tilde u_{1b}(t-x)\m
\,u_{c1}(t+x'-2\pi)\m\,\tilde u_{2d}(t-x')\,\cr
&&\quad-\,u_{c2}(t+x'-2\pi)\m\,\tilde u_{1d}(t-x')\m
\,u_{a1}(t+x-2\pi)\m\,\tilde u_{2b}(t-x)\Big)\,\,\tilde U_{21}\,\tilde U_{12}
\cr\cr
&&=\ -\m\epsilon_{be}\epsilon_{df}\,\Big(u_{a2}(t+x-2\pi)\m\,u_{e2}(t-x)\m
\,u_{c1}(t+x'-2\pi)\m\,u_{f1}(t-x')\,\cr
&&\ \qquad\qquad-\,u_{c2}(t+x'-2\pi)\m\,u_{f2}(t-x')\m
\,u_{a1}(t+x-2\pi)\m\,u_{e1}(t-x)\Big)\,\,\tilde U_{21}\,\tilde U_{12}\,,
\cr\cr
&&\quad[g_{ab}^0(t,x),\m g_{cd}^0(t,x')]\cr\cr
&&=\ \ \epsilon_{be}\epsilon_{df}\,\Big(u_{a2}(t+x-2\pi)\m\,u_{e1}(t-x)\m,
\,u_{c2}(t+x'-2\pi)\m\,u_{f1}(t-x')\,\cr
&&\ \qquad\qquad-\,u_{c2}(t+x'-2\pi)\m\,u_{f1}(t-x')\m,
\,u_{a2}(t+x-2\pi)\m\,u_{e1}(t-x)\Big)\,\,\tilde U_{22}^{^{\m\,2}}
\label{00}
\qqq
in the action on $\,\CH_{_{j_{_0}j_{_\pi}{_k\over^2}}}$.
Using the commutation relations (\ref{ur}), it is easy to show that
the expressions in the parenthesis vanish implying (\ref{4}).
\vskip 0.3cm

As for the commutators in (\ref{3}), they be may expanded in the same 
way. It will be instructive to study them for a general values of
$j$ to see what changes for the case $j={k\over^2}$.
In the action on a general component $\,\CH_{_{j_{_0}j_{_\pi}j}}\m$,
\qq
&&\quad[g^-_{ab}(t,x)\m,\,g^0_{cd}(t,x')]\ 
+\ [g^0_{ab}(t,x)\m,\,g^-_{cd}(t,x')]\cr\cr
&&=\ \epsilon_{be}\,\epsilon_{df}\,\bigg[\Big(u_{a1}(t+x-2\pi)\,\,
u_{e1}(t-x)\,\,u_{c1}(t+x'-2\pi)\,\,u_{f1}(t-x')\cr
&&\hspace{0.56cm}\qquad\qquad-\m u_{c1}(t+x'-2\pi)\,\,u_{f1}(t-x')\,\,
u_{a2}(t+x-2\pi)\,\,u_{e1}(t-x)\Big)\,\,\tilde U_{12}\,\tilde U_{22}\cr
&&\hspace{0.56cm}\qquad+\ \Big(-\m u_{a1}(t+x-2\pi)
\,\,u_{e2}(t-x)\,\,u_{c1}(t+x'-2\pi)\,\,u_{f1}(t-x')\cr
&&\hspace{0.56cm}\qquad\qquad+\m u_{c1}(t+x'-2\pi)\,\,u_{f2}(t-x')\,\,
u_{a1}(t+x-2\pi)\,\,u_{e1}(t-x)\Big)\,\,\tilde U_{11}\,\tilde U_{12}\cr
&&\hspace{0.5cm}\qquad+\ \Big(u_{a2}(t+x-2\pi)\,\,u_{e1}(t-x)\,\,
u_{c1}(t+x'-2\pi)\,\,u_{f1}(t-x')\cr
&&\hspace{0.56cm}\qquad\qquad-\m u_{c2}(t+x'-2\pi)\,\,u_{e1}(t-x')\,\,
u_{a1}(t+x-2\pi)\,\,u_{e1}(t-x)\Big)\,\,\tilde U_{22}\,\tilde U_{12}\cr
&&\hspace{0.56cm}\qquad+\ \Big(-\m u_{a1}(t+x-2\pi)\,\,u_{e1}(t-x)\,
\,u_{c1}(t+x'-2\pi)\,\,u_{f2}(t-x')\cr
&&\hspace{0.56cm}\qquad\qquad+\m u_{c1}(t+x'-2\pi)\,\,u_{f1}(t-x')\,\,
u_{a1}(t+x-2\pi)\,\,u_{e2}(t-x)\Big)\,\,\tilde U_{12}\,\tilde U_{11}\bigg].
\hspace{0.8cm}
\label{146}
\qqq
After reordering with the use of (\ref{ur}) one obtains
the expression 
\qq
\epsilon_{be}\,\epsilon_{df}\,\sum\limits_{i=1}^4X^i_{cfae}
\,\Big(T(\hat p)_{i1}\,\tilde U_{12}\m\tilde U_{22}+T(\hat p)_{i2}\,
\tilde U_{11}\m\tilde U_{12}+T(\hat p)_{i3}\,\tilde U_{22}\m \tilde U_{12}
+T(\hat p)_{i4}\,\tilde U_{12}\m\tilde U_{11}\Big),
\hspace{0.6cm}
\qqq
where
\qq
X^1_{cfae}&=&u_{c2}(t+x'-2\pi)\,\,u_{f1}(t-x')\,\,
u_{a1}(t+x-2\pi)\,\,u_{e1}(t-x)\,,\cr\cr
X^2_{cfae}&=&u_{c1}(t+x'-2\pi)\,\,u_{f2}(t-x')\,\,
u_{a1}(t+x-2\pi)\,\,u_{e1}(t-x)\,,\cr\cr
X^3_{cfae}&=&u_{c1}(t+x'-2\pi)\,\,u_{f1}(t-x')\,\,
u_{a2}(t+x-2\pi)\,\,u_{e1}(t-x)\,,\cr\cr
X^4_{cfae}&=&u_{c1}(t+x'-2\pi)\,\,u_{f1}(t-x')\,\,
u_{a1}(t+x-2\pi)\,\,u_{e2}(t-x)\,.
\qqq
For $p<k+1$, the matrix $\,T(p)\,$ is
\qq
\left(\matrix{{q^{1\pm1}[p-3]\over[p-1]}&-{q^{-p+2\pm1}[p-3]
\over[p-1][p-2]}&{q^{\mp(p-3)}\over[p-2]}-1&0\cr
\ &\ &\ &\ \cr
-{q^{p-1\pm p\mp1}([p]+[p-2])\over[p-1]^2[p]}&{q^{\pm(p-1)}
(1-[p-2]^2)\over[p-1]^2[p-2]}+1&{q^{p-2\pm1}\over[p-2]}&
-{q^{-1\mp1}[p-2]\over[p]}\cr\ &\ &\ &\ \cr
{q^{\pm p\mp1}(1-[p]^2)\over[p-1]^2[p]}-1&{q^{-p+1\pm p\mp1}
([p-2]+[p])\over[p-1]^2[p-2]}&{q^{-1\pm1}[p]\over[p-2]}&
{q^{-p\mp1}\over[p]}\cr\ &\ &\ &\ \cr
-{q^{p\mp1}[p+1]\over[p-1][p]}&-{q^{1\mp1}[p+1]\over[p-1]}&
0&{q^{\mp p\mp1}\over[p]}+1}\right).
\qqq
The general commutation relations between $\tilde U_{11},\,\tilde U_{22}$ 
and $\tilde U_{12}$ on the space $\,\CV$ of (\ref{cv}) with generic
values of $z$ may be read from (\ref{ada}). By continuity, they
still hold for $z=1$ (in $\m\tilde U_{i\ell}$, unlike in $\m U_{i\ell}$, 
there no divisions by factors that become singular at $z=1$).
They take on $\,V_{_{q^{p_{_0}}}}\otimes V_{_{q^{p_{_\pi}}}}\otimes 
V_{_{q^{p}}}\,$ the form
\qq
&\tilde U_{12}\m\tilde U_{22}\m\,q^{p-1}\,+\,\tilde U_{11}\m\tilde U_{12}
\m\,[p]\,-\,\tilde U_{12}\m \tilde U_{11}\m\,q^{-1}[p-1]\ =\ 0\,,&
\label{uu}\\\cr
&\tilde U_{12}\m \tilde U_{22}\m\,[p-2]\,-\,\tilde U_{11}\m\tilde U_{12}
\m\,q^{-p+1}\,-\,\tilde U_{22}\m\tilde U_{12}\m\,q^{-1}[p-1]\ =\ 0\,.&
\nonumber
\qqq
The vectors $\,(\tilde U_{12}\m \tilde U_{22},\,\tilde U_{11}\m
\tilde U_{12},\,\tilde U_{22}\m\tilde U_{12},\,\tilde U_{12}\m
\tilde U_{11})\,$ solving (\ref{uu}) are null-vectors of the matrix 
$\m T(p)$ assuring the vanishing of the commutator (\ref{146}).
\vskip 0.3cm

When $p=k+1$, due to the modification of the commutation relations 
(\ref{ur}) of the $u$'s (the replacement of the matrix $\m\tilde D^{\pm}$ 
by $\m\tilde D^{'\pm}$), the last column of the matrix $\,T(p)\,$ should 
be replaced by zero. Note that the last line of $\,T(p)$, \m unlike
the last column, becomes automatically zero for $p=k+1$. It is easy 
to find then the null-vectors of the resulting $3\times3$ matrix
\qq
\left(\matrix{{q^{1\pm1}[4]\over[2]}&{q^{3\pm1}[4]\over[2][3]}
&-{q^{\pm1}[4]\over[3]}\cr\ &\ &\ \cr-q^{-2\mp2}&{q^{\mp1}[4]+1\over[3]}&
-{q^{-3\pm1}\over[3]}\cr\ &\ &\ \cr-1&{q^{2\mp2}\over[3]}&
{q^{-1\pm1}\over[3]}}\right).
\qqq
They are proportional to $\,(1,\,q^{-2},\,q^2+1)\,$ and, if the locality 
is to hold, so must be the vector $\,(\tilde U_{12}\m \tilde U_{22},\,
\tilde U_{11}\m\tilde U_{12},\,\tilde U_{22}\m\tilde U_{12})$. 
\,For $\m p=k+1$, the relations (\ref{uu}) reduce to the equations
\qq
&-\m\tilde U_{12}\m\tilde U_{22}\m\,q^{-2}\,+\,\tilde U_{11}\m\tilde U_{12}
\,-\,\tilde U_{12}\m\tilde U_{11}\m\,q^{-1}\m[2]\ =\ 0\,,&\cr\cr
&\tilde U_{12}\m\tilde U_{22}\m\,[3]\,+\,\tilde U_{11}\m\tilde U_{12}
\m\,q^2\,-\,\tilde U_{22}\m\tilde U_{12}\m\,q^{-1}\m[2]\ =\ 0&
\qqq
and imply the desired relation if $\,\tilde U_{12}\m\tilde U_{11}=0\m$. 
The latter equality fails on the space $\,V_{_{q^{p_{_0}}}}\otimes 
V_{_{q^{p_{_\pi}}}}\otimes V_{_{q^{k+1}}}\,$ but it holds on
the quotient space of $\,(V_{_{q^{p_{_0}}}}\otimes 
V_{_{q^{p_{_\pi}}}}\otimes V_{_{q^{k+1}}})^{^{inv}}$ by the null
invariant tensors. Indeed, the quotient space is non-zero only 
if the fusion rule
$\,j={k\over2}\m\leq\m{\rm min}\m(j_{_0}+j_{_\pi},\,k-j_{_0}+j_{_\pi})$,
is satisfied, i.e.\,\,if $\m j_{_0}+j_{_{\pi}}={k\over2}\,$ or 
$\,p_{_0}+p_{_{\pi}}=k+2$. But then $\,\tilde U_{11}$ vanishes on
$\,(V_{_{q^{p_{_0}}}}\otimes V_{_{q^{p_{_\pi}}}}\otimes 
V_{_{q^{k+1}}})^{^{inv}}$ as follows from the explicit
formula (\ref{dme}) for its eigenvalue.
\vskip 0.4cm

The other extreme cases $j=0,{1\over 2}$ are similar (in fact, 
the commutation relations are symmetric under the simultaneous 
change of spins $\m j\mapsto{k\over2}-j\m$ and the interchange 
of the raising and lowering components of the vertex operators). 
This ends the proof of locality of the bulk fields $\m g(t,x)\m$ 
in the action on the boundary Hilbert space.
\vskip 0.6cm

\subsection{Boundary 1-point functions of the bulk fields}
\vskip 0.2cm

Given the explicit expressions for the local fields
$g(t,x)$ it is not difficult to calculate low-order
correlation functions involving these fields, in
particular, their 1-point functions given by the 
matrix elements between the states
\qq
|j,m\rangle\otimes|p_{_0},p_{\pi},p\rangle\ 
\ \in\,\ \CH_{_{j_0j_\pi}}\,,
\qqq
see (\ref{cHs0},\ref{cHs}), where $\,|j,m\rangle\,$ is the 
state in \,$\CV_{_{k,j}}$ annihilated by the positive current modes 
$\,J_n\,$ and such that $\,(J^3_0-m)\,|j,m\rangle=0\,$ 
and $\,|p_{_0},p_{_\pi},p\rangle\,$ is in the (one-dimensional space) 
$\,(\tilde\CV_{_{q^{p_{_0}}}}\otimes\tilde
\CV_{_{q^{p_{_\pi}}}}\otimes\tilde\CV_{_{q^p}})^{^{inv}}\,$ with
$\,p=2j+1$. \,The structure (\ref{cmp}) of fields $\,g\,$ reduces
this calculation to that of the matrix elements of $\,\tilde U\,$
given above by Eqs.\,\,(\ref{dme},\ref{dmep},\ref{ndme},\ref{ndmep}).
It remains then to calculate the matrix elements 
\qq
\langle j,m'\m\,|\m\,u_{ai}(t+x-2\pi)\,\m\tilde u_{\ell b}(t-x)\,\m
|\,j,m\rangle\,,
\label{utu}
\qqq
where $j'=j$ for $i=\ell=1$ or $i=\ell=2$, $j'=j+1$ for
$i=2$, $\ell=1$ and $j'=j-1$ for $i=1$, $\ell=2$. This is, 
in fact, a special conformal 4-point block whose computation
may  be found in \cite{HST}. Clearly, the expression (\ref{utu})
is time independent so that we may set $t=0$. It is more
convenient to use the exponential coordinate. \,Set $\,\ln{z}
=i(x-2\pi)$ and $\,\ln{\tilde z}=-ix$. \,The conformal block corresponds 
to insertion of the primary field labeled of spin $j$ at zero, spin
$j'$ at infinity and spin ${1\over2}$ at $z$ and $\tilde z$, 
see Fig.\,\,5.

\leavevmode\epsffile[-70 -20 255 310]{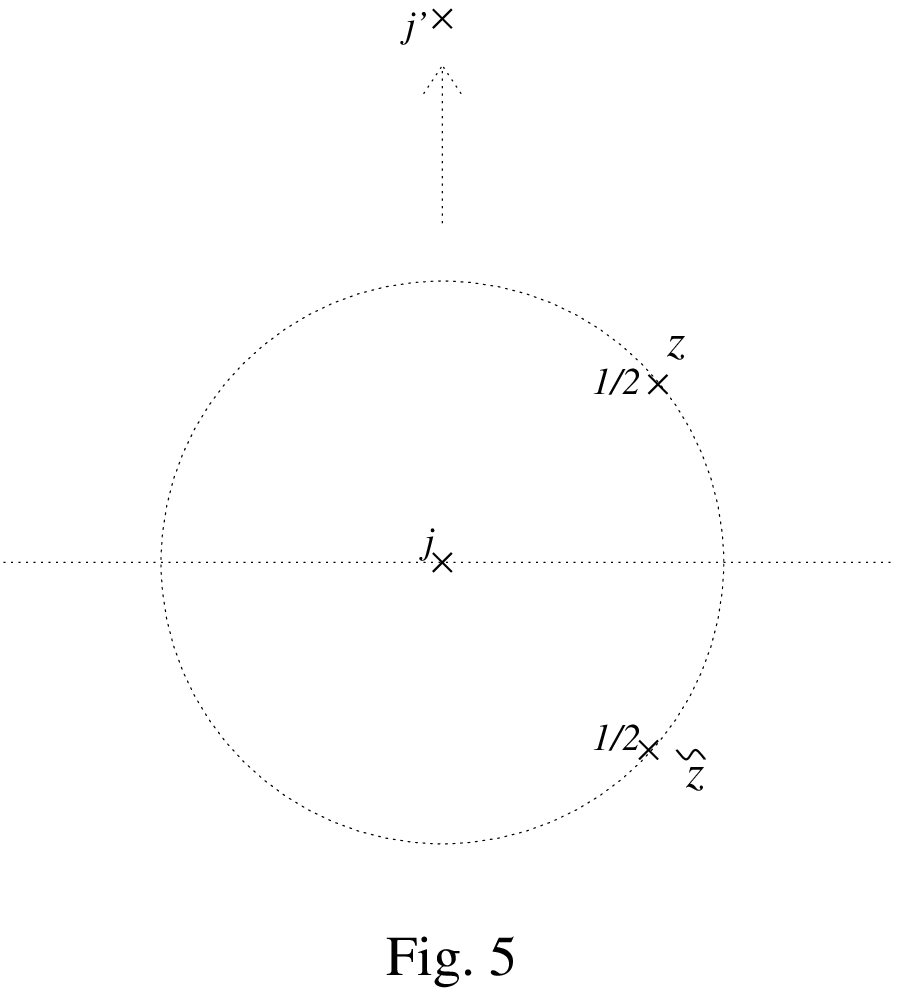}

\noindent The conformal invariance imposes the form
\qq
(iz)^{\Delta_{1/2}}\,(i{\tilde z})^{\Delta_{1/2}}\,
z^{\Delta_{j'}-2\Delta_{1/2}-\Delta_j}
\,\,F_{\,\,i\ell}^{m,a,b,m'}(\eta)
\qqq
where $\Delta_j={j(j+1)\over k+2}$ is the conformal weight
of the spin $j$ primary field and $\eta=\tilde z/z$.
\,In its dependence on $(m',a,b,m)$, $\,F_{i\ell}$ is an invariant 
tensor in the tensor product $\,\CV_{_{j'}}\otimes\CV_{_{1/2}}
\otimes\CV_{_{1/2}}\otimes\CV_{_j}\,$ of four representations 
of the $SU(2)$ group and it satisfies the Knizhnik-Zamolodchikov equation
\qq
\left[(k+2){d\over d\eta}+{C_{23}\over1-\eta}-{C_{34}\over\eta}\right]
F_{i\ell}(\eta)\ =\ 0\,,
\label{KZ}
\qqq
where $\,C_{23}$ and $\,C_{34}$ are the Casimir operator of $su(2)$ 
acting in the product of the two middle (spin $1/2$) representations 
and of the last two ones, respectively. 
\vskip 0.3cm

When $j'=j$, there are two linearly independent $SU(2)$ invariants 
$I_0$ and $I_1$ and they may be chosen so that 
\qq
&&\hbox to 4.5cm{$C_{23}\,I_0\,=\,{_1\over^2}\,I_0+\,I_1\,,$\hfill}
C_{23}\,I_1\,=\,-{_3\over^2}\,I_1\,,\cr
&&\hbox to 4.5cm{$C_{34}\,I_0\,=\,-(j+1)\,I_0\,,$\hfill}
C_{34}\,I_1\m=I_0+\,j\,I_1\,.
\label{Cs}
\qqq
Eq.\,\,(\ref{KZ}) reduces to the hypergeometric one and two solutions
corresponding to $\,F_{11}$ and $F_{22}$ are \cite{HST}
\qq
&&F_{11}(\eta)\,=\,{n_{11}(j)\,\,\eta^{j\over k+2}\over
(1-\eta)^{3\over2(k+2)}}\,\Big[\,2j
\,(1-\eta)\,\,F(1-{_1\over^{k+2}},1+{_{2j}\over^{k+2}},1+{_{2j+1}
\over^{k+2}};\,\eta)\,\,I_0\,\cr
&&\hspace{4cm}+\,(2j+1)\,\,F(-{_1\over^{k+2}},{_{2j}\over^{k+2}},{_{2j+1}
\over^{k+2}};\,\eta)\,\,I_1\,\Big]\,,\cr\cr
&&F_{22}(\eta)\,=\,{n_{22}(j)\,\,\eta^{-{j+1\over k+2}}
\over(1-\eta)^{3\over2(k+2)}}\,\Big[\,(k+1-2j)
\,(1-\eta)\,\,F(1-{_1\over^{k+2}},1-{_{2j+2}\over^{k+2}},1-{_{2j+1}
\over^{k+2}};\,\eta)\,\,I_0\,\hspace{0.8cm}\cr
&&\hspace{4cm}-\,\eta\,\,F(1-{_1\over^{k+2}},1-{_{2j+2}
\over^{k+2}},2-{_{2j+1}\over^{k+2}};\,\eta)\,\,I_1\,\Big]
\qqq
behaving when $\eta\to0$ as $\,\CO(\eta^{\Delta_{j\pm1/2}-\Delta_{1/2}
-\Delta_j})$, respectively. The commutation relations
(\ref{qrh}) with the $D$-matrices (\ref{Dpd}) fix the ratio
$\,\,{n_{22}(j)\over n_{11}(j)}=(2j+1)\,{\Gamma({2j+1\over k+2})
\,\Gamma(1-{2j\over k+2})\over \Gamma(1-{2j+1\over k+2})\,
\Gamma({2j+2\over k+2})}\,$.
\vskip 0.3cm

When $j'=j\pm1$, up to normalization, there is only one invariant 
tensor $\,J_\pm$ in the space $\,\CV_{_{j'}}
\otimes\CV_{_{1/2}}\otimes\CV_{_{1/2}}\otimes\CV_{_j}\,$
and the Kznizhnik Zamolodchikov equation takes the scalar
form
\qq
0\ =\ \left[\,{d\over d\eta}+{\Delta_1-2\Delta_{1/2}\over1-\eta}
-{\Delta_{j\pm1/2}-\Delta_{1/2}-\Delta_j\over\eta}\,\right]\,
\cases{F_{21}(\eta)\cr F_{12}(\eta)}
\qqq
with the solutions
\qq
&&F_{12}=n_{12}(j)\,\,\eta^{\Delta_{j-1/2}-\Delta_{1/2}-\Delta_j}\,
(1-\eta)^{\Delta_1-2\Delta_{1/2}}\,J_-=n_{12}(j)\,\,\eta^{-{j+1\over k+2}}
\,(1-\eta)^{1\over2(k+2)}\,J_-\,,\hspace{0.8cm}\cr
&&F_{21}=n_{21}(j)\,\,\eta^{\Delta_{j+1/2}-\Delta_{1/2}-\Delta_j}
\,(1-\eta)^{\Delta_1-2\Delta_{1/2}}\,J_+=n_{21}(j)\,\,
\eta^{j\over k+2}\,(1-\eta)^{1\over 2(k+2)}\,J_+\,.\hspace{0.8cm}
\qqq
The normalizations $\,n_{i\ell}(j)$ could be constraint further
using the explicit free field realizations of the operators $\chi$ 
and $\tilde u$.
\vskip 1cm

\noindent{\bf Acknowledgements}. The authors are greatful
to the Max-Planck Institute for Mathematics in the Sciences
in Leipzig where this work was initiated for hospitality.
K.G. and I. T. acknowledge also the support of the Erwin 
Schr\"{o}dinger International Institute for Mathematical 
Physics in Vienna and P.T. that of the International Center
for Theoretical Physics in Trieste. K.G. would like to thank 
J\"urg Fr\"ohlich for explanation of the relevance 
of Chern-Simons Wilson lines in the theory of Quantum Hall Effect. 
I.T. thanks Anton Alekseev for discussions and the Institut
des Hautes Etudes Scientifiques in Bures-sur-Yvette for 
invitation which has permitted completion of the paper. His 
work was supported in part by the Bulgarian 
National Council for Scientific Research under contract F-828. 

\vskip 0.5cm

\nappendix{1}
\vskip 0.6cm

We prove here that the monodromy operator $\,\tilde U\,$ commutes 
with the co-product action of $\,\CU_q(su(2))$ in the space (\ref{ds}).
We have to show that
\qq
\Delta\gamma_{_1}\m\,\tilde U_{_2}
\ =\ \tilde U_{_2}\m\,\Delta\gamma_{_1}\,.
\qqq
This is follows with the use of the commutation relations (\ref{gpm}) 
and (\ref{aga}). Indeed,
\qq
&&(\gamma_{_{0-}}\gamma_{_{\pi-}}\gamma_{_w}\,
\gamma_{_{\pi+}}^{-1}\m\gamma_{_{0+}}^{-1})_{_1}\,\,
(a\,\m\gamma_{_{\pi-}}^{-1}\m\gamma_{_{\pi+}}\m\tilde a)_{_2}\cr\cr
&&=\ (\gamma_{_{0-}}\gamma_{_{\pi-}})_{_1}
(\gamma_{_w})_{_1}\,a_{_2}(\gamma_{_{\pi+}}^{-1})_{_1}
(\gamma_{_{\pi-}}^{-1})_{_2}\,(\gamma_{_{\pi+}}\m\tilde a)_{_2}
\,(\gamma_{_{0+}}^{-1})_{_1}\cr\cr
&&=\ (\gamma_{_{0-}}\gamma_{_{\pi-}})_{_1}\,a_{_2}\m(R^-)^{-1}\,
(\gamma_{_w})_{_1}\,(\gamma_{_{\pi-}}^{-1})_{_2}\,(\gamma_{_{\pi+}}^{-1})_{_1}
\m R^+\,(\gamma_{_{\pi+}})_{_2}\,\tilde a_{_2}\,(\gamma_{_{0+}}^{-1})_{_1}
\cr\cr
&&=\ a_{_2}\,(\gamma_{_{0-}})_{_1}\,(\gamma_{_{\pi-}})_{_1}
\m(R^-)^{-1}\,(\gamma_{_{\pi-}}^{-1})_{_2}\,(\gamma_{_w})_{_1}
(\gamma_{_{\pi+}})_{_2}\m R^+\,(\gamma_{_{\pi+}}^{-1})_{_1}
\,\tilde a_{_2}\,(\gamma_{_{0+}}^{-1})_{_1}
\cr\cr
&&=\ a_{_2}\,(\gamma_{_{0-}})_{_1}\,(\gamma_{_{\pi-}}^{-1})_{_2}
\m(R^-)^{-1}\,(\gamma_{_{\pi-}})_{_1}\,(\gamma_{_{\pi+}})_{_2}
\,(\gamma_{_w})_{_1}\m R^+\,\tilde a_{_2}\,(\gamma_{_{\pi+}}^{-1}\m
\gamma_{_{0+}}^{-1})_{_1}
\cr\cr
&&=\ (a\,\m\gamma_{_{\pi-}}^{-1})_{_2}\,(\gamma_{_{0-}})_{_1}\,
(\gamma_{_{\pi+}})_{_2}\,(\gamma_{_{\pi-}})_{_1}\,\tilde a_{_2}\,
(\gamma_{_w}\m\gamma_{_{\pi+}}^{-1}\m\gamma_{_{0+}}^{-1})_{_1}\cr\cr
&&=\ (a\,\m\gamma_{_{\pi-}}^{-1}\m\gamma_{_{\pi+}}\m
\tilde a_{_2})_{_2}\,\,(\gamma_{_{0-}}\m\gamma_{_{\pi-}}\m
\gamma_{_w}\m\gamma_{_{\pi+}}^{-1}\m\gamma_{_{0+}}^{-1})_{_1}\,.
\qqq
\vskip 0.5cm

\nappendix{2}
\vskip 0.6cm

We prove here the commutation relation (\ref{ada}).
\qq
&&U_{_1}\,(D^+)^{-1}\, U_{_2}\ =\ (b\m\,\gamma_{_{\pi-}}^{-1}\,
\gamma_{_{\pi+}}\,\tilde a)_{_1}\,(D^+)^{-1}\,(b\m\,
\gamma_{_{\pi-}}^{-1}\,\gamma_{_{\pi+}}\,\tilde a)_{_2}\cr\cr
&&=\ (b\m\,\gamma_{_{\pi-}}^{-1}\,\gamma_{_{\pi+}})_{_1}
\,b_{_2}\,(R^+)^{-1}\,\tilde a_{_1}\,(\gamma_{_{\pi-}}^{-1}\,
\gamma_{_{\pi+}}\,\tilde a)_{_2}\cr\cr
&&=\ b_{_1}\,b_{_2}\,(\gamma_{_{\pi-}}^{-1}\,\gamma_{_{\pi+}})_{_1}
\,(R^+)^{-1}\,(\gamma_{_{\pi-}}^{-1}\,\gamma_{_{\pi+}})_{_2}\,
\tilde a_{_1}\,\tilde a_{_2}\cr\cr
&&=\ (D^\pm)^{-1}\,(b_{_2}\,b_{_1}\, R^\pm\,(\gamma_{_{\pi-}}^{-1})_{_1}
\,(\gamma_{_{\pi-}}^{-1})_{_2}\,(R^+)^{-1}\,(\gamma_{_{\pi+}})_{_1}\,
(\gamma_{_{\pi+}})_{_2}\, R^\mp\,\tilde a_{_2}\,\tilde a_{_1}\,
(D^\mp)^{-1}\cr\cr
&&=\ (D^\pm)^{-1}\,b_{_2}\,b_{_1}\,(\gamma_{_{\pi-}}^{-1})_{_2}\,
(\gamma_{_{\pi-}}^{-1})_{_1}\,R^-\,(\gamma_{_{\pi+}})_{_2}\,
(\gamma_{_{\pi+}})_{_1}\,\tilde a_{_2}\,\tilde a_{_1}\,
(D^\mp)^{-1}\cr\cr
&&=\ (D^\pm)^{-1}\,(b\m\,\gamma_{_{\pi-}}^{-1})_{_2}\,b_{_1}\,
(\gamma_{_{\pi+}})_{_2}\,R^-\,(\gamma_{_{\pi-}}^{-1})_{_1}\,
\tilde a_{_2}\,(\gamma_{_{\pi+}}\,\tilde a)_{_1}\,(D^\mp)^{-1}\cr\cr
&&=\ (D^\pm)^{-1}\,(b\m\,\gamma_{_{\pi-}}^{-1}\gamma_{_{\pi+}})_{_2}\,
b_{_1}\, R^-\,\tilde a_{_2}\,(\gamma_{_{\pi-}}^{-1}\,\gamma_{_{\pi+}}\,
\tilde a)_{_1}\,(D^\mp)^{-1}\cr\cr
&&=\ (D^\pm)^{-1}\,(b\m\,\gamma_{_{\pi-}}^{-1}\gamma_{_{\pi+}}\,
\tilde a)_{_2}\,D^-\,(b\,\gamma_{_{\pi-}}^{-1}\,\gamma_{_{\pi+}}\,
\tilde a)_{_1}\,(D^\mp)^{-1}\ =\ (D^\pm)^{-1}\,U_{_2}\,D^-\,U_{_1}
\,(D^\mp)^{-1}\,.
\nonumber
\qqq

\end{document}